\documentclass[prb,twocolumn,showpacs]{revtex4}
\usepackage{color,soul}
\usepackage{amsmath}
\usepackage{dcolumn}
\usepackage{graphicx}
\usepackage{bm}
\usepackage{amssymb}
\usepackage{subfigure}
\begin{document}

\title{Interplay between antiferromagnetic spin fluctuation and electron-phonon coupling and the origin of the peak-dip-hump structure in the anti-nodal spectrum of high-$T_{c}$ cuprate superconductors}
\author{Xinyue Liu and Tao Li}
\email{litao_phys@ruc.edu.cn}
\affiliation{School of Physics, Renmin University of China, Beijing 100872, P.R.China}

\begin{abstract}
Electron-phonon coupling is believed to be responsible for many spectral anomalies in the cuprate superconductors. In particular, the $B_{1g}$ buckling mode of the oxygen ion in the $CuO_{2}$ plane has been proposed to be responsible for the dramatic peak-dip-hump(PDH) structure in the anti-nodal spectrum. The recent observation of the exceptional flat quasiparticle dispersion in the anti-nodal region and the sudden suppression of the PDH structure around the pseudogap end point cast doubts on such a scenario. Instead, a scenario involving the coupling to the antiferromagnetic spin fluctuation seems to resolve both puzzles naturally. Here we present a systematic study on the interplay between antiferromagnetic spin fluctuation and electron-phonon coupling in the cuprate superconductors and their respective roles in the origin of the PDH structure. We show that the coupling strength to the $B_{1g}$ buckling mode is strongly renormalized by the vertex correction caused by the antiferromagnetic spin fluctuation. In particular, the coupling strength to the $B_{1g}$ buckling mode is strongly suppressed in the $\mathbf{q}\rightarrow 0$ limit as a result of the destructive interference between electron-phonon coupling at electron momentum differ by the antiferromagnetic wave vector. Counterintuitively, we find that the same vertex correction enhances the phonon contribution to the PDH structure as a result of its highly anisotropic momentum dependence around $\mathbf{q}=0$. We also find that while the coupling to either the antiferromagnetic spin fluctuation or the $B_{1g}$ buckling mode can generate a PDH structure in the anti-nodal spectrum with similar phenomenologies, the sudden suppression of such a structure around the pseudogap end point should be mainly attributed to the dramatic change in the nature of the spin fluctuation at such a critical doping. We suggest to take the PDH structure in the anti-nodal spectrum as a spectral signature for the emergence of fluctuating local moment in the pseudogap phase and the entrance of a doped Mott insulating state.          
\end{abstract}

\maketitle

\section{Introduction}
The electron-phonon coupling is believed to play an important role in shaping the cuprate physics. The earliest evidence in this respect is provided by the unexpected broadness of the quasiparticle peak in the antiferromagnetic parent compound and lightly doped systems, a signature which has been attributed to the polaronic electron-phonon coupling in the system\cite{Damascelli,KMShen,Nagaosa1}. In the more heavily doped cuprates, coupling to different phonon modes have been invoked to interpret various spectral anomalies observed in the nodal and the anti-nodal region. More specifically, the half-breathing Cu-O bond stretching mode at 70 meV is proposed to be responsible for the kink anomaly in the nodal direction\cite{Damascelli,Zhou}. On the other hand, the out-of-plane $B_{1g}$ oxygen buckling phonon at 35 meV is proposed to be responsible for the peak-dip-hump(PDH) structure in the anti-nodal region\cite{He,Zhou}. 

Beside these spectral anomalies, electron-phonon coupling is also believed to play an important role in enhancing the d-wave pairing which is dominantly mediated by the antiferromagnetic spin fluctuation. In particular, a recent ARPES measurement made on a quasi-one dimensional cuprates indicates that an additional attractive interaction between electron residing on nearest neighboring sites must be invoked to explain the shape of the electron spectrum. The appearance of such an attractive interaction is rather unusual in a Mott insulating background of the cuprate system and is most likely induced by Su-Schrieffer-Hegger(SSH)-type electron-phonon coupling\cite{attractive1,attractive2,attractive3}. Indeed, a recent quantum Monte Carlo study shows that the effective interaction mediated by SSH-type electron-phonon coupling can not only enhance electron pairing in the d-wave channel, but can also enhance the tendency toward itinerant antiferromagnetic order\cite{Yao1,Yao2,Yao3}. More generally, electron-phonon coupling is argued to be involved in the intertwinement between different competing orders in the pseudogap phase of the cuprate phase diagram\cite{CDW}. Last but not least, phonon may provide the ultimate sink for both energy and momentum relaxation in the realization  of the strange metallic transport behavior of the cuprates\cite{Cooper,Taillefer,Proust,Hussey,Phillips,Hartnoll}.

While the electron-phonon coupling is playing such an important role in the cuprate physics, elucidating its exact role remains a challenging open question as a result of the strongly correlated nature of the cuprate superconductors\cite{RLiu,Ginsberg,TPD,Gunnarsson,Nagaosa,Shen,Peng}. On the theoretical side, first principle calculation indicates that the electron-phonon coupling strength in the cuprate superconductors is order of magnitude smaller than that needed to interpret the observed spectral anomalies in ARPES measurements\cite{Louie,Heid}. It is argued that such a discrepancy may be attributed to the strong correlation effect neglected in the first principle calculation\cite{Reznik}. However, exactly how the electron correlation effect would affect the electron-phonon coupling in the cuprate superconductors is not clear. Naively, the suppression of the electron itinerancy by the strong correlation effect would suppress the Coulomb screening in the system and thus enhance the bare electron-phonon coupling strength with underdoping. On the other hand, the suppression of the charge fluctuation near the Mott insulating phase may also render the electron-phonon coupling irrelevant at low energy\cite{Nagaosa}. 

As a result of these difficulties, most previous studies on the effect of the electron-phonon coupling in the cuprate superconductors are limited at a rather qualitative level. A particular example in this respect is the lasting debate on the origin of the PDH structure in the anti-nodal spectrum.\cite{Fink1,Gromko,Kim,Dessau,He,Chen,Zhou} While it is now widely believed that the coupling to the $B_{1g}$ oxygen buckling mode plays an important role in the origin of such a spectral anomaly, many puzzles remains. For example, it is observed in ARPES measurement that the quasiparticle peak exhibits an exceptionally flat dispersion around the anti-nodal point in the superconducting state, so flat that is totally beyond our expectation for the Bogoliubov quasiparticle in a standard d-wave BCS superconductor\cite{Dessau,He,Chen}. At the same time, such a sharp quasiparticle peak is always accompanied by a rather broad spectral continuum at higher binding energy, so broad that is beyond our expectation for any given Boson mode. It is thus highly skeptical if the coupling to the $B_{1g}$ oxygen buckling mode could account for all these extraordinary spectral features considering the almost local nature of this phonon mode. 

A more severe challenge on the electron-phonon coupling scenario comes from the peculiar doping dependence of the PDH structure as exposed by recent ARPES measurements. More specifically, it is found that the strength of the PDH structure suffers a sudden suppression around the pseudogap end point\cite{He,Chen}, where quantum critical behavior of unknown origin has been identified in recent years in various measurements\cite{Proust,Hussey}. In the electron-phonon coupling scenario, such a sudden suppression in the strength of the PDH structure would imply a corresponding sudden drop in the electron-phonon coupling strength around the pseudogap endpoint\cite{He}. However, even if we admit such a possibility, we must admit that the interplay between the electron-phonon coupling and the strong correlation effect is playing a crucial role in the generation of PDH structure. The reason is rather clear - the majority of experimental evidences point to the fact that the pseudogap phenomena in the cuprate superconductors is a genuine strong correlation effect and that the electron-phonon coupling is playing at most a secondary role in its origin. 

The very coincidence of the sudden suppression of the PDH structure and the vanishing of the pseudogap phenomena suggests an alternative way to understand the origin of such a spectral anomaly. More specifically, it is reasonable to take the PDH structure as the spectral signature for the entrance of the pseudogap phase, a doped Mott insulating state driven by strong correlation effect between the electrons. In such a scenario, it is the strong correlation effect, rather than the electron-phonon coupling that makes the dominant contribution to the PDH structure.   

The most direct consequence of the electron correlation effect in the pseudogap phase is the emergence of antiferromagnetic fluctuation of local moment. The existence of such fluctuating local moment has been well documented by the extensive RIXS measurements conducted in the last decades\cite{Tacon,Dean}. More recently, it is found that there is a dramatic change in the nature of spin fluctuation from itinerant particle-hole-continuum-like to local moment-like around the pseudogap endpoint\cite{Minola,Zhu}. Such fluctuating local moment can have profound effect on the cuprate physics. Beside being a major candidate for the pairing glue of the d-wave pairing, the antiferromagnetic spin fluctuation is also generally believed to be responsible for the various non-fermi liquid behavior in the strange metal and the pseudogap phase of the cuprate phase diagram. However, a rigorous treatment of such fluctuating local moment and its coupling with the electron is a difficult problem. At the phenomenological level, a widely adopted approach to describe the local moment physics in the cuprate superconductors is to resort to the so called spin-fermion model, in which the itinerant quasiparticle excitation and the local moment fluctuation are treated as two independent degree of freedom\cite{MMP,Chubukov,Liu}. 

A recent study based on the spin-fermion model indicates that the coupling to the antiferromagnetic spin fluctuation can induce dramatic spectral anomalies in the anti-nodal region\cite{Li}. In particular, it is found that the anti-nodal quasiparticle can acquire an extremely flat dispersion as a result of its coupling to the antiferromagnetic spin fluctuation, which scatters the quasiparticle between different anti-nodal regions in the Brillouin zone. In addition, the broadness of the high energy hump structure also find a natural explanation from the diffusive nature of the local moment fluctuation. In this scenario, the sudden suppression of the PDH structure around the pseudogap end point should be attributed to the dramatic change in the nature of spin fluctuation from local-moment-like to itinerant-particle-hole-continuum-like occurs there. 

According to such a scenario, far from being a standard d-wave BCS Bogliubov quasiparticle, the sharp quasiparticle peak found in the anti-nodal spectrum should be understood as a composite object made up of the Bogliubov quasiparticle and intense dressing cloud of antiferromagnetic spin fluctuation. The strong non-BCS character of the anti-nodal quasiparticle excitation implies that it would also experience phonon scattering in a way very different from that of the bare BCS Bogliubov quasiparticle. Thus before claiming the dominance of the spin fluctuation contribution to the PDH structure, it is important to understand how would the electron-phonon coupling contribute and interplay with the antiferromagnetic spin fluctuation in the generation of such a spectral anomaly. As we discussed above, a study of such interplay effect is equally important even if we assume a dominant phonon contribution to the PDH structure.       
        
With these considerations in mind, we present in this work a systematic perturbative study of the interplay between the electron-phonon coupling and the antiferromagnetic spin fluctuation effect in the cuprate superconductors. We find that the coupling to either the antiferromagnetic spin fluctuation or the $B_{1g}$ buckling phonon mode can generate a PDH structure in the anti-nodal spectrum with very similar phenomenology. At the same time, we find that the coupling to the $B_{1g}$ buckling mode suffers strong renormalization from the vertex correction by the antiferromagnetic spin fluctuation. In particular, we find that the coupling strength to the $B_{1g}$ buckling mode is strongly suppressed in the $\mathbf{q}=0$ limit as the result of destructive interference effect from such a vertex correction. Surprisingly, we find that the same vertex correction enhances the phonon contribution to the PDH structure as a result of the singular momentum dependence of the vertex correction around $\mathbf{q}=0$. Our result indicates that both the antiferromagnetic spin fluctuation and the $B_{1g}$ buckling mode act cooperatively to generate the PDH structure in the anti-nodal spectrum. 

Our result indicates that the transmutation of the nature of spin fluctuation is the main driving force for the observed sudden suppression of the PDH structure around the pseudogap endpoint. No sudden change in the electron-phonon coupling strength is necessary to account for the unusual doping dependence of the PDH structure. The PDH structure in the anti-nodal spectrum can thus be taken as a sharp spectral signature for the entrance of the doped Mott insulating phase or the emergence of local moment fluctuation below the doping level corresponding to the pseudogap endpoint.

The paper is organized as follows. In the next section we will introduce the phenomenological spin-fermion model supplemented with electron-phonon coupling to the $B_{1g}$ buckling mode. A detailed analysis of the momentum dependence of the electron-phonon coupling vertex will be the focus of this section. In the third section, we will analyze the self-energy correction from the antiferromagnetic spin fluctuation and the electron-phonon coupling separately to see if either of these two scenario can account for the observed PDH structure in the anti-nodal spectrum. In the fourth section, we will introduce the vertex correction to the electron-phonon coupling by the antiferromagnetic spin fluctuation. A detailed analysis of the momentum structure in the vertex correction at the first order will be presented in this section. We will also present and solve numerically a self-consistent equation for the full vertex function based on the ladder approximation. From this analysis, the intricate interplay between the electron-phonon coupling and the antiferromagnetic spin fluctuation is clearly exposed. In the fifth section, we calculate the electron spectral function in the anti-nodal region with the effect of vertex correction included. The cooperative nature of the antiferromagnetic spin fluctuation and the electron-phonon coupling to the $B_{1g}$ buckling mode can be seen directly from the result of such a tedious calculation. In the last section, we conclude our result and discuss other consequence of the interplay between the antiferromagnetic spin fluctuation and the electron-phonon coupling in the cuprate superconductors.

\section{A spin-fermion model of the superconducting cuprates with SSH-type coupling to the $B_{1g}$ buckling mode}
\subsection{Model Hamiltonian}
In this work, we model superconducting cuprates with an extended spin-fermion model. The electronic part of the model Hamiltonian is given by
\begin{equation}
H=H_{BCS}+H_{sf}+H_{SSH}
\end{equation}
Here
\begin{eqnarray}
H_{BCS}&=&\sum_{\mathbf{k},\alpha}\epsilon_{\mathbf{k}}c^{\dagger}_{\mathbf{k},\alpha}c_{\mathbf{k},\alpha}\nonumber\\
&+&\sum_{\mathbf{k}}\Delta_{\mathbf{k}}(c^{\dagger}_{\mathbf{k},\uparrow}c^{\dagger}_{-\mathbf{k},\downarrow}+c_{-\mathbf{k}\downarrow}c_{\mathbf{k},\uparrow})
\end{eqnarray}
in which 
\begin{equation}
\epsilon_{\mathbf{k}}=-2t(\cos k_{x}+\cos k_{y})-4t^{'}\cos k_{x}\cos k_{y}-\mu
\end{equation}
and
\begin{equation}
\Delta_{\mathbf{k}}=\frac{\Delta}{2}(\cos k_{x}-\cos k_{y})
\end{equation}
are the kinetic energy and the d-wave pairing potential of the itinerant quasiparticle in the cuprates. 

To simulate the electron correlation effect in the cupates, we assume that these quasiparticles are coupled to antiferromagnetic spin fluctuation of the local moment in the form\cite{Chubukov}
\begin{equation}
H_{sf}=g\sum_{i}\mathbf{S}_{i}\cdot \mathbf{s}_{i}
\end{equation} 
in which $\mathbf{S}_{i}$ denotes the fluctuating local moment at site $i$ and
\begin{equation} 
\mathbf{s}_{i}=\frac{1}{2}\sum_{\alpha,\beta}c^{\dagger}_{i,\alpha}\bm{\sigma}_{\alpha,\beta}c_{i,\beta}
\end{equation}
is the spin density operator of the quasiparticle at site $i$. Such a coupling can also be derived from a mean field decoupling of the interaction term in the Hubbard model. Here we assume that the antiferromagnetic fluctuation of the local moment is described by a Gaussian action of the form
\begin{equation}
\mathcal{S}_{eff}=\frac{1}{\beta} \sum_{\mathbf{q},i\omega_{n}}\chi^{-1}(\mathbf{q},i\omega_{n})\mathbf{S}(\mathbf{q},i\omega_{n})\cdot \mathbf{S}(-\mathbf{q},-i\omega_{n})
\end{equation}
in which $\chi(\mathbf{q},i\omega_{n})$ denotes the generalized susceptibility of the local moment. A widely used form for $\chi(\mathbf{q},i\omega_{n})$ is the Millis-Monien-Pines(MMP) susceptibility, which reads\cite{MMP,SWZ}
\begin{equation}
\chi(\mathbf{q},i\omega_{n})=\frac{\chi_{0}}{1+\xi^{2}(\mathbf{q}-\mathbf{Q})^{2}+\frac{|\omega_{n}|}{\omega_{sf}}+\frac{\omega^{2}_{n}}{\Delta^{2}_{s}}}
\end{equation}
here $\omega_{sf}$ denotes the characteristic frequency of the Landau damped local moment fluctuation, $\xi$ denotes the correlation length of the local moment. $\chi_{0}$ is the static spin susceptibility of the local moment at the antiferromagnetic wave vector. The last term in the denominator, $\frac{\omega^{2}_{n}}{\Delta^{2}_{s}}$, which is usually ignored in the MMP susceptibility, is introduced here to ensure convergence of total spin fluctuation spectral weight. $\Delta_{s}$ can be interpreted as the spin gap of the local moment fluctuation when the Landau damping term is turn off. As will be clear below, the overall scale for the coupling between the electron and the local moment is set by the product $g^{2}\chi_{0}$.

The electron-phonon coupling with the oxygen buckling mode in the CuO$_{2}$ plane take the form of Su-Schrieffer-Hegger(SSH) coupling\cite{Nagaosa,Shen} and reads
\begin{equation}
H_{SSH}=\lambda\sum_{i,\bm{\delta},\sigma}u_{i,\bm{\delta}}( c^{\dagger}_{i+\bm{\delta},\sigma}c_{i,\sigma}+h.c.)
\end{equation} 
here $u_{i,\bm{\delta}}$ denotes the phonon coordinate of the oxygen buckling mode on the bond connecting site $i$ and its nearest neighboring site $i+\bm{\delta}$. $\bm{\delta}=x,y$. $\lambda$ is the bare electron-phonon coupling strength. We note that such a linear coupling is only permitted when the position of the oxygen ion is away from the inversion center. In the momentum space, the SSH coupling Hamiltonian takes the form of 
\begin{equation}
H_{SSH}=\frac{1}{\sqrt{N}}\sum_{\mathbf{k,q},\sigma,\bm{\delta}}F_{\bm{\delta}}(\mathbf{k,q})u_{\mathbf{q},\bm{\delta}}c^{\dagger}_{\mathbf{k+q},\sigma}c_{\mathbf{k},\sigma}
\end{equation}
Here $u_{\mathbf{q},\bm{\delta}}$ and $c_{\mathbf{k},\sigma}$ are the Fourier component of the phonon coordinate and the electron operator and are given by
\begin{eqnarray}
u_{i,\bm{\delta}}&=&\frac{1}{\sqrt{N}}\sum_{i}u_{\mathbf{q},\bm{\delta}}e^{i\mathbf{q}\cdot\mathbf{r}_{i}}\nonumber\\
c_{i,\sigma}&=&\frac{1}{\sqrt{N}}\sum_{i}c_{\mathbf{k},\sigma}e^{i\mathbf{q}\cdot\mathbf{r}_{i}}
\end{eqnarray}
$F_{\bm{\delta}}(\mathbf{k,q})$ is the bare electron-phonon vertex in the momentum space and is given by
\begin{equation}
F_{\bm{\delta}}(\mathbf{k,q})=\lambda[e^{i\mathbf{k}\cdot \bm{\delta}}+e^{-i\mathbf{(k+q)}\cdot \bm{\delta}}]
\end{equation}
$N$ denotes the number of lattice site. In this study, we will focus on the coupling to the oxygen buckling mode in the $B_{1g}$ channel. For this purpose, we decompose the phonon coordinate $u_{\mathbf{q},\bm{\delta}}$ as follows
\begin{eqnarray}
u_{\mathbf{q},x}&=&\frac{1}{\sqrt{2}}(u^{A}_{\mathbf{q}}+u^{B}_{\mathbf{q}})\nonumber\\
u_{\mathbf{q},y}&=&\frac{1}{\sqrt{2}}(u^{A}_{\mathbf{q}}-u^{B}_{\mathbf{q}})
\end{eqnarray} 
in which $u^{A}_{\mathbf{q}}$ and $u^{B}_{\mathbf{q}}$ denote the phonon coordinate in the $A_{1g}$ and $B_{1g}$ channel. The electron-phonon coupling in the $B_{1g}$ channel then reads
\begin{equation}
H_{SSH}=\frac{1}{\sqrt{N}}\sum_{\mathbf{k,q},\sigma}f(\mathbf{k,q})u^{B}_{\mathbf{q}}c^{\dagger}_{\mathbf{k+q},\sigma}c_{\mathbf{k},\sigma}
\end{equation}
with
\begin{equation}
f(\mathbf{k,q})=\frac{\lambda}{\sqrt{2}}\sum_{\bm{\delta}}s_{\bm{\delta}}(e^{i\mathbf{k}\cdot\bm{\delta}}+e^{-i\mathbf{(k+q)}\cdot\bm{\delta}})
\end{equation}
Here $s_{x,y}=\pm1$ denotes the phase factor of the $B_{1g}$ mode. For simplicity we assume that the $B_{1g}$ buckling mode is dispersionless and has a frequency of $\Omega$. 

An analysis of the momentum dependence of the electron-phonon coupling vertex $f(\mathbf{k,q})$ is especially illuminating for the following discussion. Since $\bm{\delta}$ connects sites in different sublattices of the square lattice, we have
\begin{equation}
f(\mathbf{k,q})=-f(\mathbf{k+Q,q})
\end{equation}
for any $\mathbf{k}$ and $\mathbf{q}$. Here $\mathbf{Q}=(\pi,\pi)$ is the wave vector for antiferromagnetic ordering. As we will see below, such a non-trivial momentum dependence in $f(\mathbf{k,q})$ is responsible for a destructive interference effect in the vertex correction to the electron-phonon coupling by the antiferromagnetic spin fluctuation in the $\mathbf{q}=0$ limit. On the other hand, at the anti-nodal momentum of $\mathbf{k}=(0,\pi)$ or $(\pi,0)$, the electron-phonon coupling constant strength $|f(\mathbf{k,q})|$ reaches its maximum of $2\sqrt{2}\lambda$ at $\mathbf{q}=(0,0)$(see Fig.1 for an illustration of the momentum dependence of the electron-phonon coupling strength). We thus expect that the interplay between the antiferromagnetic spin fluctuation and the electron-phonon coupling to the $B_{1g}$ buckling mode to be most dramatic in the anti-nodal region. More generally, it can be shown that $|f(\mathbf{k,q})|$ reaches its maximum of $2\sqrt{2}\lambda$ when the following condition is met
\begin{eqnarray}
k_{x}-k_{y}&=&\pi\ mod(2\pi)\nonumber\\
2k_{x}+q_{x}&=&0\ mod(2\pi)\nonumber\\
q_{x}-q_{y}&=&0\ mod(2\pi)
\end{eqnarray}

 \begin{figure}
\includegraphics[width=9cm]{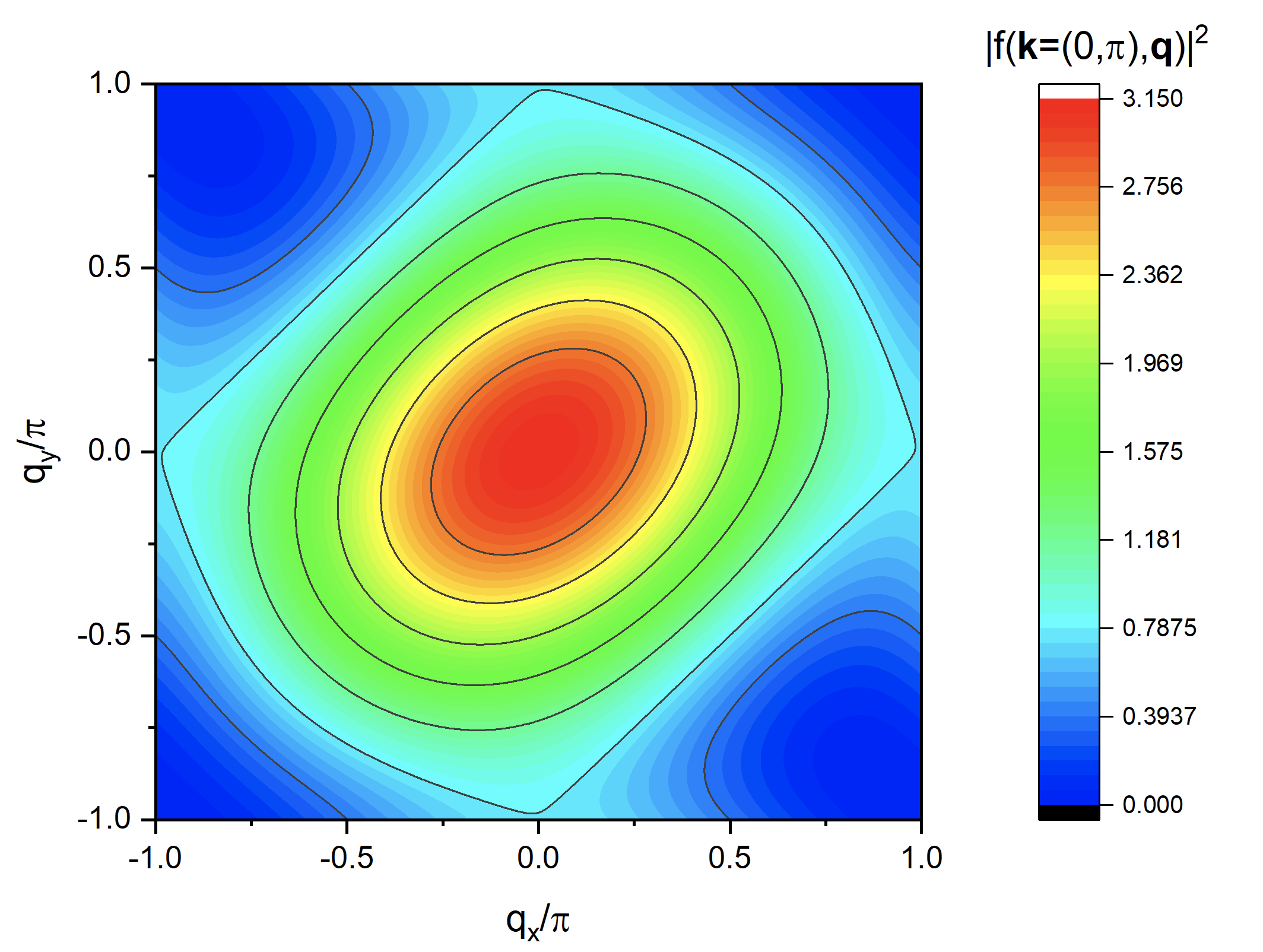}
\caption{The momentum dependence of the electron-phonon coupling strength $|f(\mathbf{k,q})|$ in unit of $\lambda$. Here we have fixed the electron momentum at one of the anti-nodal point in the Brillouin zone, $\mathbf{k}=(0,\pi)$.}
\end{figure}

 \subsection{Model parameters}
In this study, we set $t=250\ meV$ in the bare electron dispersion and use it as the unit of energy. We set $t'=-0.3t$ and $\mu=-t$. This corresponds to a system with a hole doping level of about $x=15\%$ and a hole-like fermi surface. We set $\Delta=0.12t=30\ meV$ for the d-wave pairing gap\cite{Li} and $\xi=3$ and $\omega_{sf}=0.06t=15\ meV$ for the antiferromagnetic fluctuation of the local moment. These values are typical for optimally doped cuprate superconductors\cite{Zha,Zhu}. The spin gap parameter $\Delta_{s}$ is less crucial in the calculation. Here we will set it to be $\Delta_{s}=2\Delta= 60\ meV$. With the value of $\xi$ and $\omega_{sf}$ fixed, we can estimate $\chi_{0}$ from the local spin sum rule, which requires
\begin{equation}
\langle \mathbf{S}^{2}_{i}\rangle=\frac{1}{2\pi N}\sum_{\mathbf{q}}\int_{0}^{\infty}d\omega \coth(\frac{\beta\omega}{2})R(\mathbf{q},\omega)=\frac{3}{4}(1-x)
\end{equation}
in which
\begin{equation}
R(\mathbf{q},\omega)=-2 Im \chi(\mathbf{q},\omega+i0^{+})
\end{equation}
is the spectral weight of the local moment fluctuation. Completing the above integration we find that $\chi_{0}\simeq400\ eV^{-1}=100/t$. Such a value is very consistent with that estimated from NMR data\cite{Zha}. Lastly, we set $\Omega=0.15t=37.5 \ meV$ for the $B_{1g}$ buckling mode\cite{Nagaosa,Shen,Peng}. 

Unlike the above model parameters, the coupling strength $g$ between the spin fluctuation and the electron and the electron-phonon coupling constant $\lambda$ are much more difficult to estimate. We will treat both of them as tunable parameters. In our calculation we will set both $g$ and $\lambda$ to be of order $t$.

 \section{Electron self-energy in the superconducting state and the peak-dip-hump structure in the anti-nodal spectrum}
In this section, we compute the electron self-energy caused by the scattering from the antiferromagnetic spin fluctuation and the $B_{1g}$ oxygen buckling mode separately in the d-wave superconducting state. The purpose is to see if either of these two scenarios can account for the main phenomenologies of the PDH structure in the anti-nodal region, namely, the extremely flat dispersion of the quasiparticle peak and the accompanying very broad high energy hump. We note that the calculation for the antiferromagnetic spin fluctuation has been done in our previous work and the first subsection in this section serves mainly as a brief summary of the known results and to set the stage for the upcoming discussions. 
 
 \subsection{Self-energy caused by the scattering from the antiferromagnetic spin fluctuation}
 To calculate the electron self-energy in the superconducting state, we adopt the Nambu formalism, in which the electron Green's function is defined as
 \begin{equation}
G(\mathbf{k},\tau)=-\langle T_{\tau}\psi_{\mathbf{k}}(\tau)\psi^{\dagger}_{\mathbf{k}}(0) \rangle
 \end{equation}
 here
 \begin{equation}
 \psi_{\mathbf{k}}=\left(\begin{array}{c}c_{\mathbf{k},\uparrow} \\c^{\dagger}_{-\mathbf{k},\downarrow}\end{array}\right)
 \end{equation}
 is Nambu spinor. To the second order in the spin-fermion coupling $g$, the electron self-energy caused by the scattering from the antiferromagnetic spin fluctuation is given by
 \begin{equation}
 \Sigma_{AF}(\mathbf{k},i\nu_{n})=\frac{3g^{2}}{4N\beta}\sum_{\mathbf{q},i\omega_{m}}G^{(0)}(\mathbf{k-q},i\nu_{n}-i\omega_{m})\chi(\mathbf{q},i\omega_{m})
 \end{equation}
 here
 \begin{equation}
 G^{(0)}(\mathbf{k},i\nu_{n})=\frac{1}{i\nu_{n}-\epsilon_{\mathbf{k}}\sigma_{3}-\Delta_{\mathbf{k}}\sigma_{1}}
 \end{equation}
 is the bare electron Green's function in the BCS superconducting state, $\sigma_{1}$ ad $\sigma_{3}$ are the Pauli matrices in the Nambu space. The self-energy can be decomposed into three components as follows
 \begin{equation}
 \Sigma_{AF}(\mathbf{k},i\nu_{n})= \Sigma^{(0)}(\mathbf{k},i\nu_{n})+ \Sigma^{(3)}(\mathbf{k},i\nu_{n})\sigma_{3}+ \Sigma^{(1)}(\mathbf{k},i\nu_{n})\sigma_{1}\nonumber\\
 \end{equation}  
 with
 \begin{eqnarray}
 \Sigma_{AF}^{(0)}(\mathbf{k},i\nu_{n})&=&\frac{3g^{2}}{4N\beta}\sum_{\mathbf{q},i\omega_{m}}\frac{i\nu_{n}-i\omega_{m}}{(i\nu_{n}-i\omega_{m})^{2}-E^{2}_{\mathbf{k-q}}}\chi(\mathbf{q},i\omega_{m})\nonumber\\
 \Sigma_{AF}^{(3)}(\mathbf{k},i\nu_{n})&=&\frac{3g^{2}}{4N\beta}\sum_{\mathbf{q},i\omega_{m}}\frac{\epsilon_{\mathbf{k-q}}}{(i\nu_{n}-i\omega_{m})^{2}-E^{2}_{\mathbf{k-q}}}\chi(\mathbf{q},i\omega_{m})\nonumber\\
 \Sigma_{AF}^{(1)}(\mathbf{k},i\nu_{n})&=&\frac{3g^{2}}{4N\beta}\sum_{\mathbf{q},i\omega_{m}}\frac{\Delta_{\mathbf{k-q}}}{(i\nu_{n}-i\omega_{m})^{2}-E^{2}_{\mathbf{k-q}}} \chi(\mathbf{q},i\omega_{m})\nonumber\\
 \end{eqnarray}
 Inserting the spectral representation of $\chi(\mathbf{q},i\omega_{m})$, namely
 \begin{equation}
 \chi(\mathbf{q},i\omega_{m})=\frac{1}{2\pi}\int d\omega\frac{R(\mathbf{q},\omega)}{i\omega_{m}-\omega}
 \end{equation} 
 and completing the summation over the Mastubara frequency, we get the imaginary part of the retarded electron self-energy as follows
 \begin{eqnarray}
-Im \Sigma_{AF}^{(0)}(\mathbf{k},\omega+i0^{+})&=&\frac{3g^{2}}{16N}\sum_{\mathbf{q},s=\pm1} K(\mathbf{k,q},s,\omega)\nonumber\\
-Im \Sigma_{AF}^{(3)}(\mathbf{k},\omega+i0^{+})&=&-\frac{3g^{2}}{16N}\sum_{\mathbf{q},s=\pm1} \frac{s\epsilon_{\mathbf{k-q}}}{E_{\mathbf{k-q}}}K(\mathbf{k,q},s,\omega)\nonumber\\
-Im \Sigma_{AF}^{(1)}(\mathbf{k},\omega+i0^{+})&=&-\frac{3g^{2}}{16N}\sum_{\mathbf{q},s=\pm1} \frac{s\Delta_{\mathbf{k-q}}}{E_{\mathbf{k-q}}}K(\mathbf{k,q},s,\omega)\nonumber\\
 \end{eqnarray}
 in which
 \begin{equation}
 K(\mathbf{k,q},s,\omega)=R(\mathbf{q},\omega+sE_{\mathbf{k-q}})\times[f(sE_{\mathbf{k-q}})+n_{B}(\omega+sE_{\mathbf{k-q}})]\nonumber
 \end{equation}
 here $f(E)$ and $n_{B}(E)$ are the Fermi and Bose distribution function. To account for the suppression of Landau damping effect in the superconducting state, we have cutoff the spin fluctuation spectral weight $R(\mathbf{q},\omega)$ derived from the MMP susceptibility below $\mathrm{min}_{\mathbf{k}}(\Delta_{\mathbf{k}}+\Delta_{\mathbf{k+q}})$ in the calculation.
 
 \begin{figure}
\includegraphics[width=9cm]{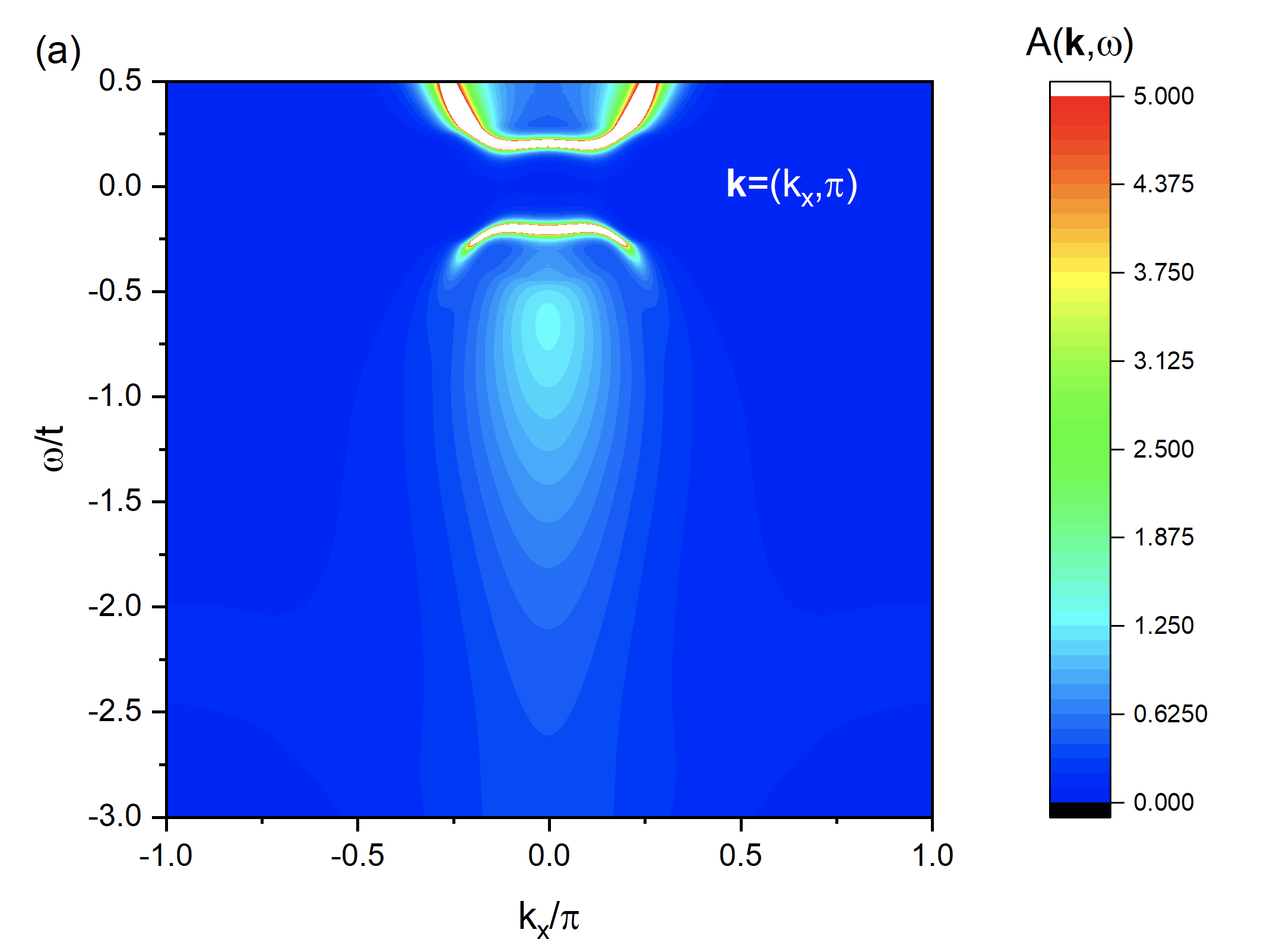}
\includegraphics[width=8cm]{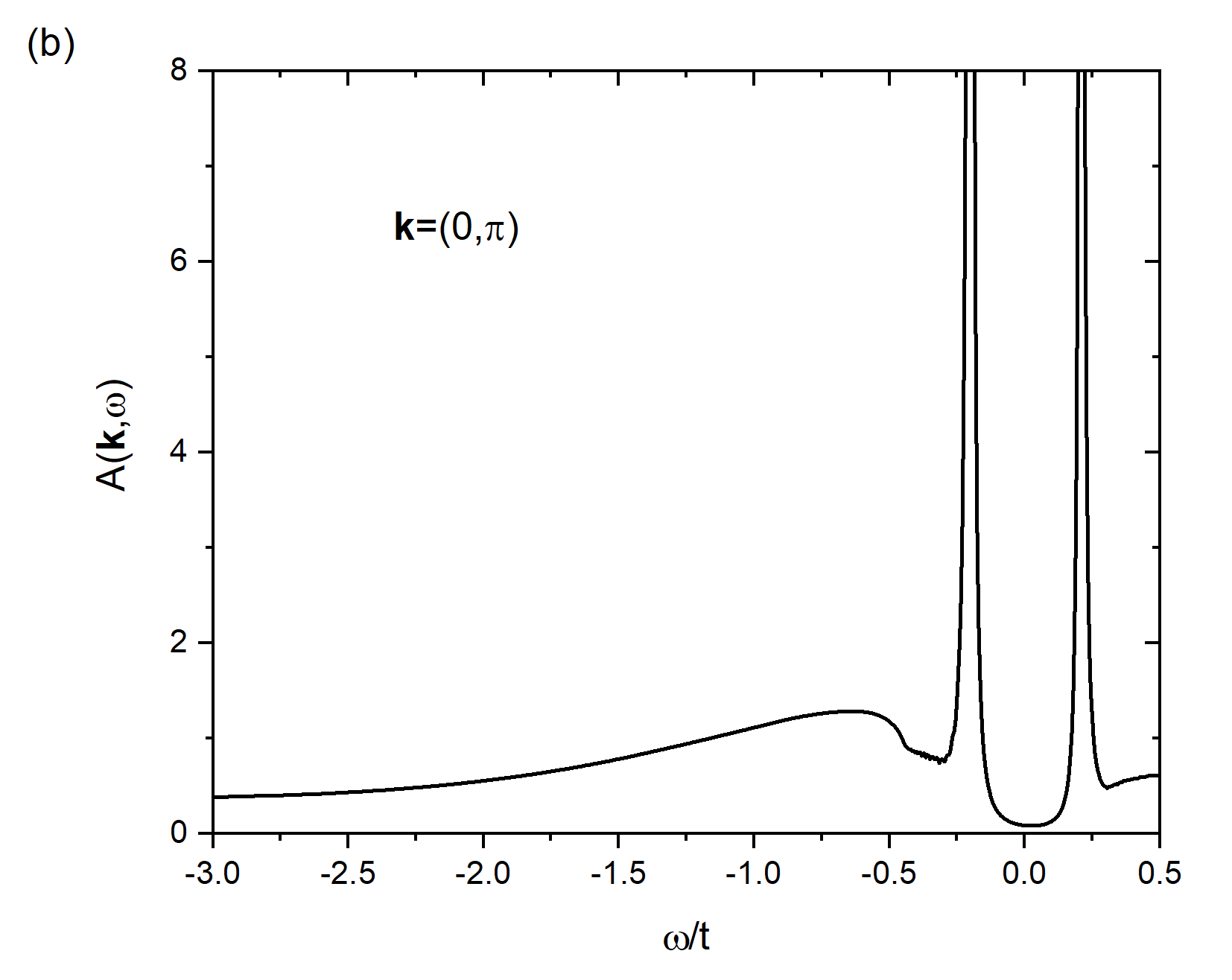}
\caption{The electron spectral function around the anti-nodal point calculated with the self-energy correction $\Sigma_{AF}$ contributed by the scattering from the antiferromagnetic spin fluctuation. (a)Distribution of the electron spectral weight along  the $(-\pi,\pi)-(\pi,\pi)$ line in the superconducting state. (b) The energy distribution curve(EDC) at the anti-nodal point $\mathbf{k}=(0,\pi)$.}
\end{figure}  
 \begin{figure}
\includegraphics[width=9cm]{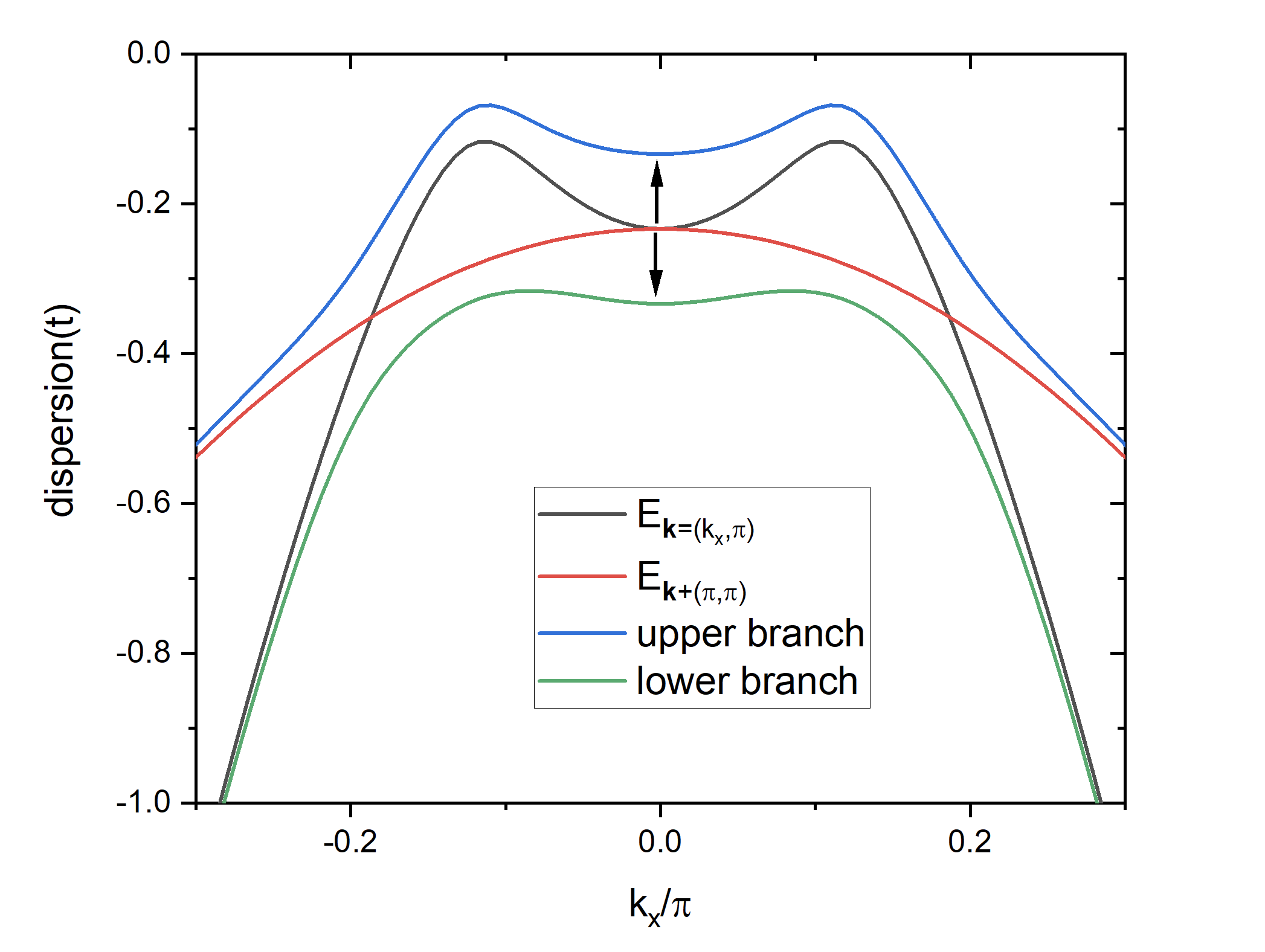}
\caption{Illustration of the level repulsion effect between the anti-nodal quasiparticle dispersion and that of the scattering final state by antiferromagnetic spin fluctuation.}
\end{figure}  

 With the electron self-energy we can compute the electron spectral function as follows
 \begin{eqnarray}
 A(\mathbf{k},\omega) &=&-2Im G_{1,1}(\mathbf{k},\omega+i0^{+})\nonumber\\
 \end{eqnarray}
 in which the subscript in $G_{1,1}$ denotes the $(1,1)$ component of the matrix Green's function. Here we will focus on the electron spectral function in the anti-nodal region. Fig.2 illustrates the spectral function calculated along the $(-\pi,\pi)-(\pi,\pi)$ line and the energy distribution curve(EDC) calculated at $\mathbf{k}=(0,\pi)$ with the model parameters specified in the last section. A clear peak-dip-hump structure is seen around the anti-nodal point. The dispersion of the quasiparticle peak is found to be extremely flat and the accompanying high energy hump structure is found to be rather broad. The extremely flat dispersion of the quasiparticle peak can be understood as the result of the level repulsion effect illustrated in Fig.3. More specifically, the scattering from the antiferromagnetic spin fluctuation couples the quasiparticle around one anti-nodal point $M=(\pi,0)$ to that around the other anti-nodal point $M'=(0,\pi)$, which are degenerate with each other at $M$ ad $M'$. The level repulsion effect induced by such a coupling will push the upper branch of the quasiparticle dispersion to lower binding energy and thus enhance its coherence. This explain the emergence of a sharp quasiparticle peak with extremely flat dispersion around the anti-nodal point.  At the same time, the lower branch of the quasiparticle dispersion will be pushed to higher binding energy. The corresponding spectral weight will be distributed into a broad continuum as a result of the diffusive nature of the local moment fluctuation. Thus the emergence and the basic characteristic of the PDH structure in the anti-nodal region can be naturally understood in the spin fluctuation scenario. To further check this understanding, we have deliberately masked the self-energy contribution from those scattering final states that has a momentum located in the shaded area as illustrated in Fig.4a. The corresponding electron spectral function in the anti-nodal region is shown in Fig.4b and Fig.4c. Clearly, the PDH structure is strongly suppressed.

 \begin{figure}
\includegraphics[width=9cm]{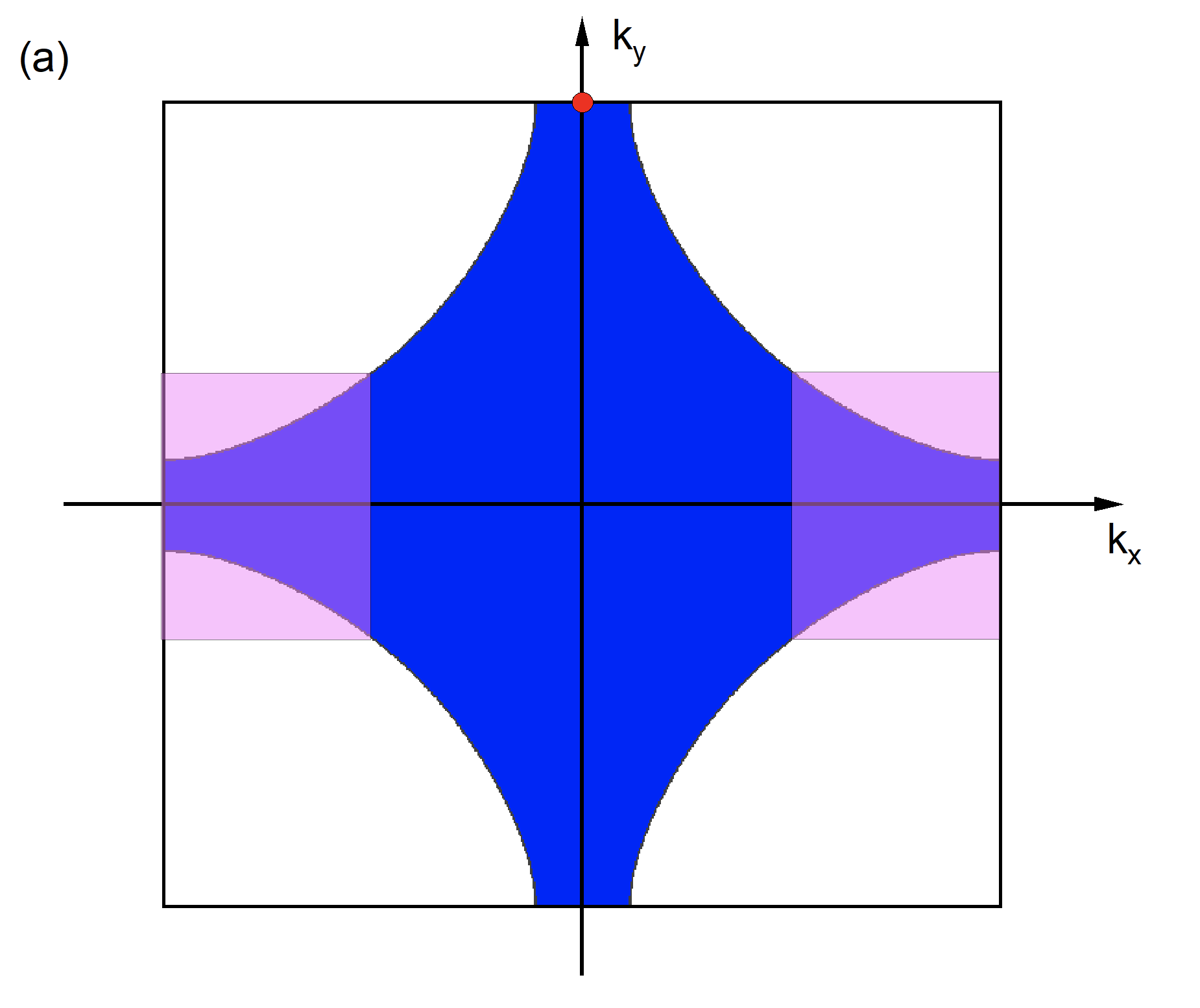}
\includegraphics[width=9cm]{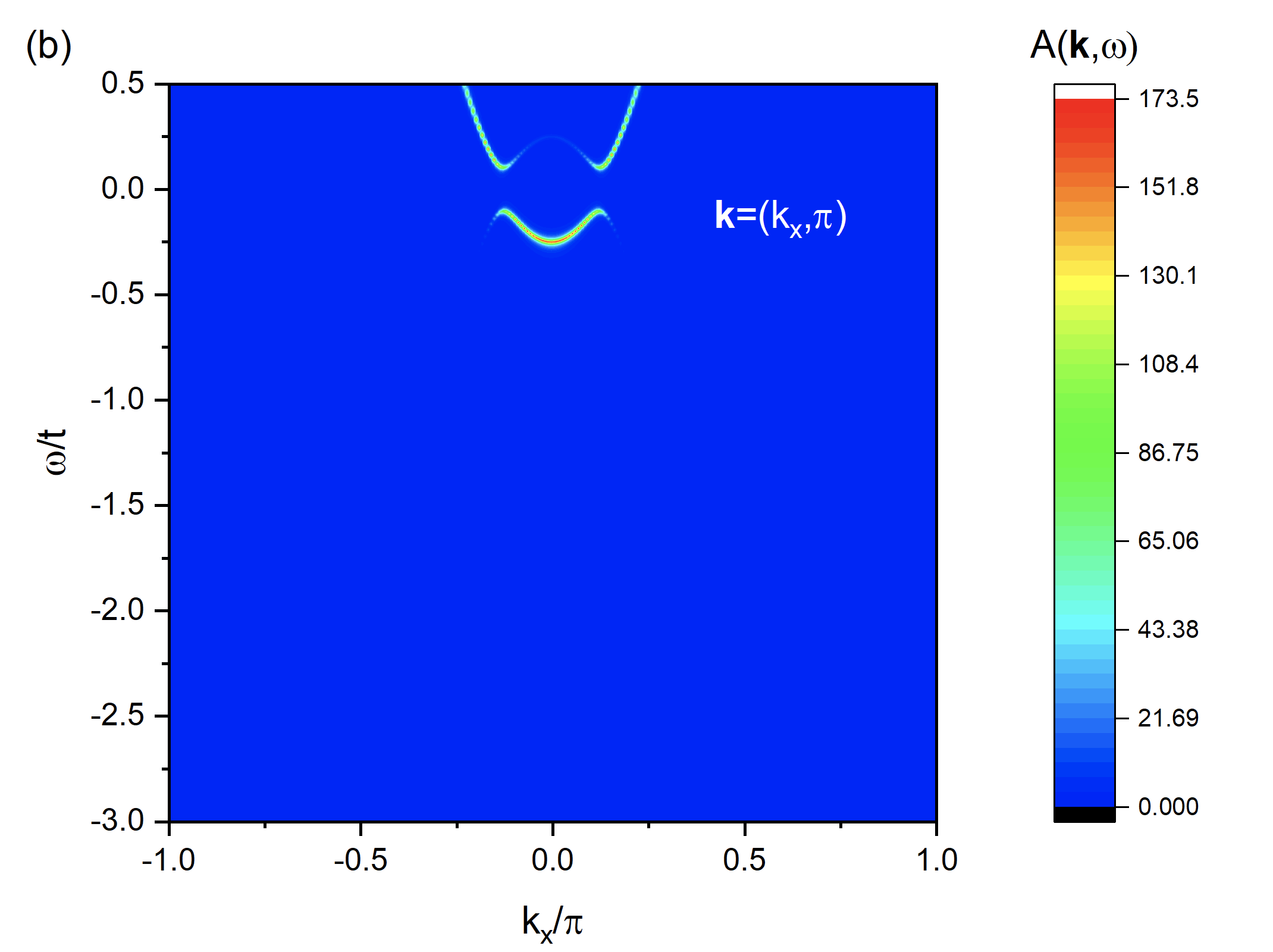}
\includegraphics[width=9cm]{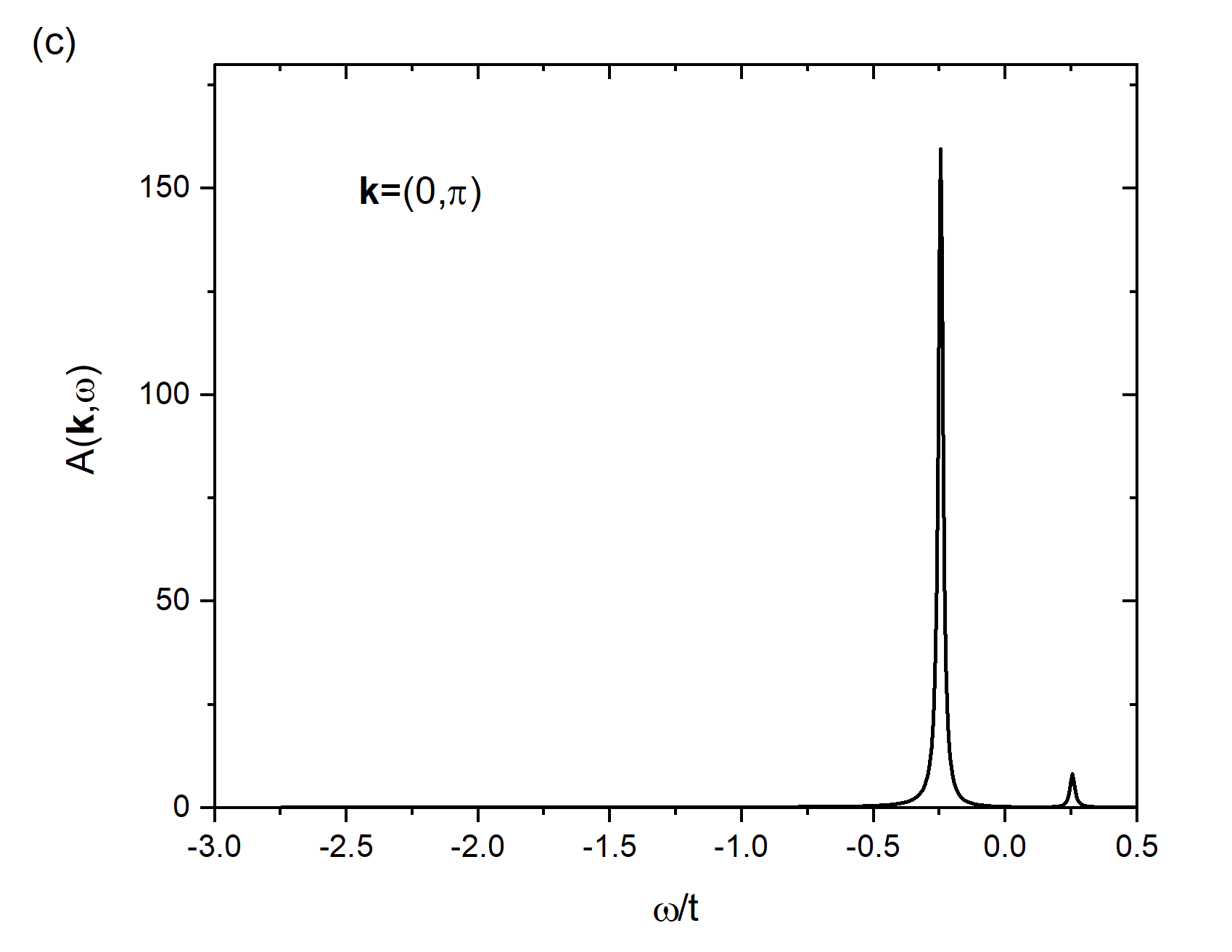}
\caption{The electron spectral function around the anti-nodal point calculated from $\Sigma_{AF}$ with the dominate momentum region of scattering final state, namely the momentum region around the other anti-nodal point in the Brillouin zone, deliberately masked. (a)The shaded area illustrated the dominate momentum region of the scattering final state in the scattering process induced by antiferromagnetic spin fluctuation. (b)Distribution of the electron spectral weight along the $(-\pi,\pi)-(\pi,\pi)$ line in the superconducting state with contribution from the shaded area masked. (c) The energy distribution curve(EDC) at the anti-nodal point $\mathbf{k}=(0,\pi)$ with contribution from the shaded area masked. The red dot denotes the position of the anti-nodal point around which the electron spectral function is calculated.}
\end{figure}  

By the way, we note that the paring gap of the renormalized quasiparticle is bigger than the pairing gap $\Delta$ that we start with. This is caused by the anomalous self-energy correction $\Sigma_{AF}^{(1)}(\mathbf{k},i\nu_{n})$ from the antiferromagnetic spin fluctuation, which mediate electron pairing in the d-wave channel. If we assume that such an anomalous self-energy has already been included in the BCS mean field Hamiltonian, we should then ignore it in the calculation to avoid double counting. For more details, see Ref.[\onlinecite{Li}].     
 
 We note that weak remnant of the PDH structure has been observed in the anti-nodal spectrum even above T$_{c}$ in the underdoped cuprate superconductors\cite{Hashimoto}. As argued by the authors of Ref.[\onlinecite{Hashimoto}], such remnant should be attributed to fluctuation effect in the pairing channel. Indeed, we find that the PDH structure disappears immediately when we set $\Delta=0$. Plotted in Fig.5 is the electron spectrum calculated along the $(-\pi,\pi)-(\pi,\pi)$ when we set $\Delta=0$. As can be seen from the figure, the main effect of the coupling to the antiferromagnetic spin fluctuation is to shift the main band upward in the anti-nodal region. However, the corresponding inward shift in the fermi momentum along the $(-\pi,\pi)-(\pi,\pi)$ line is inconsistent with the way in which the pseudogap is observed to emerge in the anti-nodal region\cite{Hashimoto2}. In fact, while the coupling to the antiferromagnetic spin fluctuation can indeed generate a "pseudogap" at the hot spot(the crossing point of the underlaying fermi surface and the boundary of the antiferromagnetic Brillouin zone), this is generally not the case at the anti-nodal point. As argued in Ref.[\onlinecite{pairing}], to account for the observed sudden emergence of the pseudogap in the anti-nodal spectrum without seeing the inward shifting in the fermi momentum along the $(-\pi,\pi)-(\pi,\pi)$ line, electron pairing of some sort must be invoked.    
   
 \begin{figure}
\includegraphics[width=9cm]{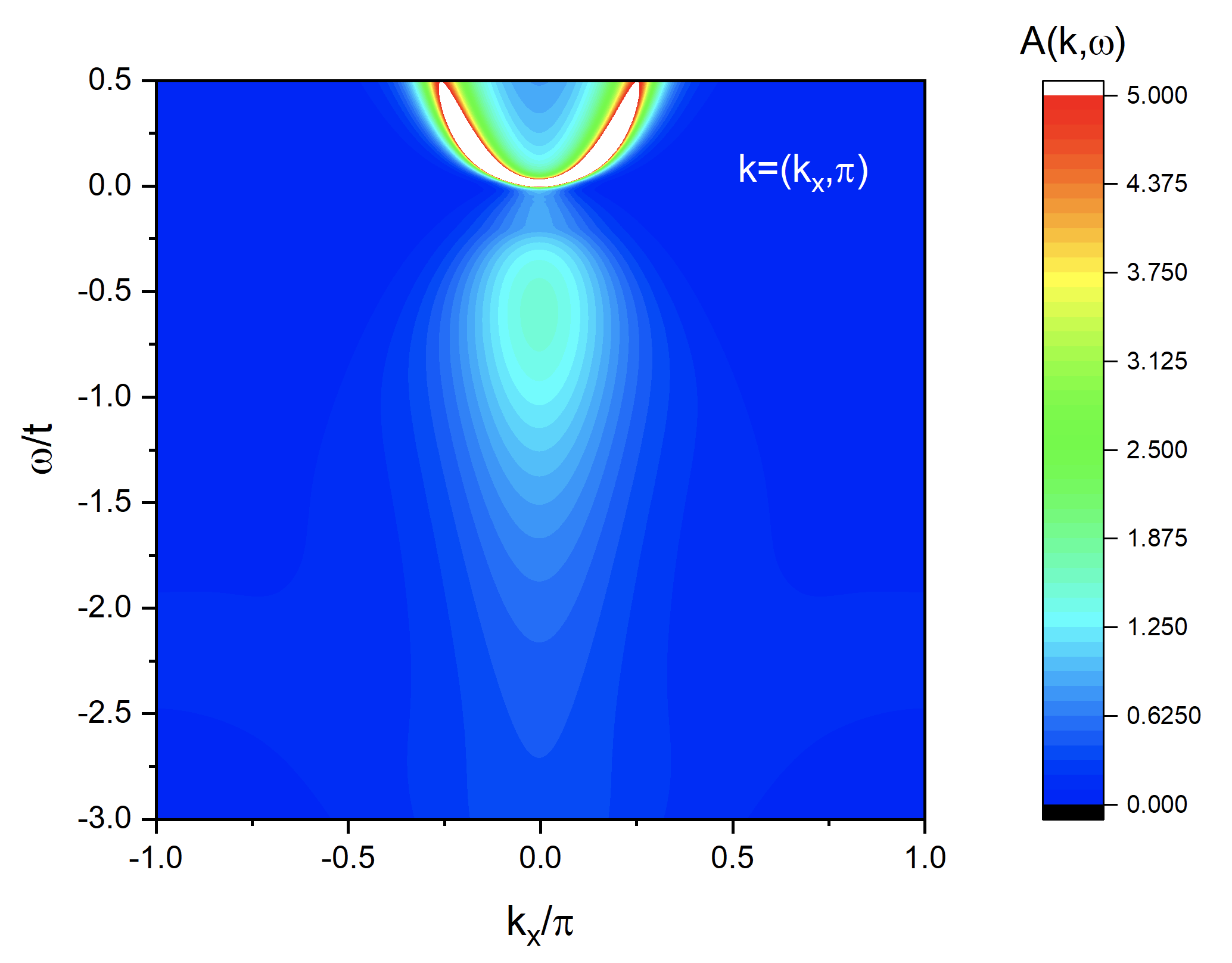}
\caption{The electron spectral function around the anti-nodal point calculated from $\Sigma_{AF}$ with the superconducting gap $\Delta$ set to zero. The PDH structure is seen to vanish identically. In the absence of the superconducting gap, the main effect of the coupling to the antiferromagnetic spin fluctuation is to shift the main band upward in the anti-nodal region. However, the corresponding inward shift in the fermi momentum along the $(-\pi,\pi)-(\pi,\pi)$ line is inconsistent with the ARPES observation. As argued in Ref.[\onlinecite{pairing}], to account for the observed sudden emergence of the pseudogap in the anti-nodal spectrum without seeing the inward shifting in the fermi momentum along the $(-\pi,\pi)-(\pi,\pi)$ line, electron pairing of some sort must be invoked.}
\end{figure}

\subsection{Self-energy from electron-phonon coupling to the $B_{1g}$ oxygen buckling mode}
Before claiming the success of the antiferromagnetic spin fluctuation scenario in the origin of the PDH structure, it is important to know if similar phenomenology can also emerge in the electron-phonon coupling scenario. Naively, this seems unlikely since the $B_{1g}$ buckling mode is essentially a local mode and there lacks the special phonon momentum needed to generate a dramatic self-energy effect around the anti-nodal point. The dispersionless nature of the $B_{1g}$ buckling mode is also seemingly at odds with the emergence of a broad hump structure. However, the following calculation indicates that the coupling to the $B_{1g}$ buckling mode can indeed generate a PDH structure in the anti-nodal spectrum with very similar phenomenology as that generated by the coupling to the antiferromagnetic spin fluctuation. Here the non-trivial momentum dependence of the electron-phonon matrix element plays a crucial role. 
      
 In the Nambu representation, the electron-phonon coupling to the $B_{1g}$ buckling mode can be written as
 \begin{equation}
 H_{SSH}=\frac{1}{\sqrt{N}}\sum_{\mathbf{k,q}}f(\mathbf{k,q})u^{B}_{\mathbf{q}}\psi^{\dagger}_{\mathbf{k+q}}\sigma_{3}\psi_{\mathbf{q}}
 \end{equation} 
 here $\sigma_{3}$ is the Pauli matrix in the Nambu space. To the second order in $\lambda$ the electron self-energy due to electron-phonon coupling is given by
 \begin{eqnarray}
 \Sigma_{PH}(\mathbf{k},i\nu_{n})&=&-\frac{1}{N\beta}\sum_{\mathbf{q},i\omega_{m}}|f(\mathbf{k,q})|^{2}\nonumber\\
 &\times&D(\mathbf{q},i\omega_{m})\sigma_{3}G^{(0)}(\mathbf{k+q},i\nu_{n}+i\omega_{m})\sigma_{3}\nonumber\\
 \end{eqnarray}
 here
 \begin{equation}
 D(\mathbf{q},i\omega_{m})=\frac{-2\Omega}{\omega^{2}_{m}+\Omega^{2}}
 \end{equation}
 is the propagator of the $B_{1g}$ buckling mode. As in the last section, the self-energy can be decomposed into three components, namely
 \begin{equation}
 \Sigma_{PH}(\mathbf{k},i\nu_{n})= \Sigma^{(0)}(\mathbf{k},i\nu_{n})+ \Sigma^{(3)}(\mathbf{k},i\nu_{n})\sigma_{3}+ \Sigma^{(1)}(\mathbf{k},i\nu_{n})\sigma_{1}\nonumber\\
 \end{equation}  
 with
 \begin{eqnarray}
 \Sigma_{PH}^{(0)}(\mathbf{k},i\nu_{n})&=&\frac{1}{N\beta}\sum_{\mathbf{q},i\omega_{m}}\frac{i\nu_{n}-i\omega_{m}}{(i\nu_{n}-i\omega_{m})^{2}-E^{2}_{\mathbf{k-q}}}\frac{2\Omega}{\omega^{2}_{m}+\Omega^{2}}\nonumber\\
 \Sigma_{PH}^{(3)}(\mathbf{k},i\nu_{n})&=&\frac{1}{N\beta}\sum_{\mathbf{q},i\omega_{m}}\frac{\epsilon_{\mathbf{k-q}}}{(i\nu_{n}-i\omega_{m})^{2}-E^{2}_{\mathbf{k-q}}}\frac{2\Omega}{\omega^{2}_{m}+\Omega^{2}}\nonumber\\
 \Sigma_{PH}^{(1)}(\mathbf{k},i\nu_{n})&=&-\frac{1}{N\beta}\sum_{\mathbf{q},i\omega_{m}}\frac{\Delta_{\mathbf{k-q}}}{(i\nu_{n}-i\omega_{m})^{2}-E^{2}_{\mathbf{k-q}}} \frac{2\Omega}{\omega^{2}_{m}+\Omega^{2}}\nonumber\\
 \end{eqnarray}
 Completing the summation over the Mastubara frequency, we get the imaginary part of the retarded electron self-energy as follows
 \begin{eqnarray}
-Im \Sigma_{PH}^{(0)}(\mathbf{k},\omega+i0^{+})&=&\frac{\pi}{2N}\sum_{\mathbf{q},s=\pm1}K'(\mathbf{k,q},s,\omega)\nonumber\\
-Im \Sigma_{PH}^{(3)}(\mathbf{k},\omega+i0^{+})&=&\frac{\pi}{2N}\sum_{\mathbf{q},s=\pm1}\frac{s\epsilon_{\mathbf{k-q}}}{E_{\mathbf{k-q}}}K''(\mathbf{k,q},s,\omega)\nonumber\\
-Im \Sigma_{PH}^{(1)}(\mathbf{k},\omega+i0^{+})&=&-\frac{\pi}{2N}\sum_{\mathbf{q},s=\pm1}\frac{s\Delta_{\mathbf{k-q}}}{E_{\mathbf{k-q}}}K''(\mathbf{k,q},s,\omega)\nonumber\\
 \end{eqnarray}
 in which
 \begin{eqnarray}
 K'(\mathbf{k,q},s,\omega)&=& | f(\mathbf{k,q})|^{2} [f(sE_{\mathbf{k+q}})+n_{B}(\Omega)] \nonumber\\
&\times& [\delta (\omega+\Omega-sE_{\mathbf{k+q}})+\delta (\omega-\Omega+sE_{\mathbf{k+q}})]\nonumber\\
K''(\mathbf{k,q},s,\omega)&=& | f(\mathbf{k,q})|^{2} [f(sE_{\mathbf{k+q}})+n_{B}(\Omega)] \nonumber\\
&\times& [\delta (\omega+\Omega-sE_{\mathbf{k+q}})-\delta (\omega-\Omega+sE_{\mathbf{k+q}})]\nonumber\\
 \end{eqnarray}

 \begin{figure}
\includegraphics[width=9cm]{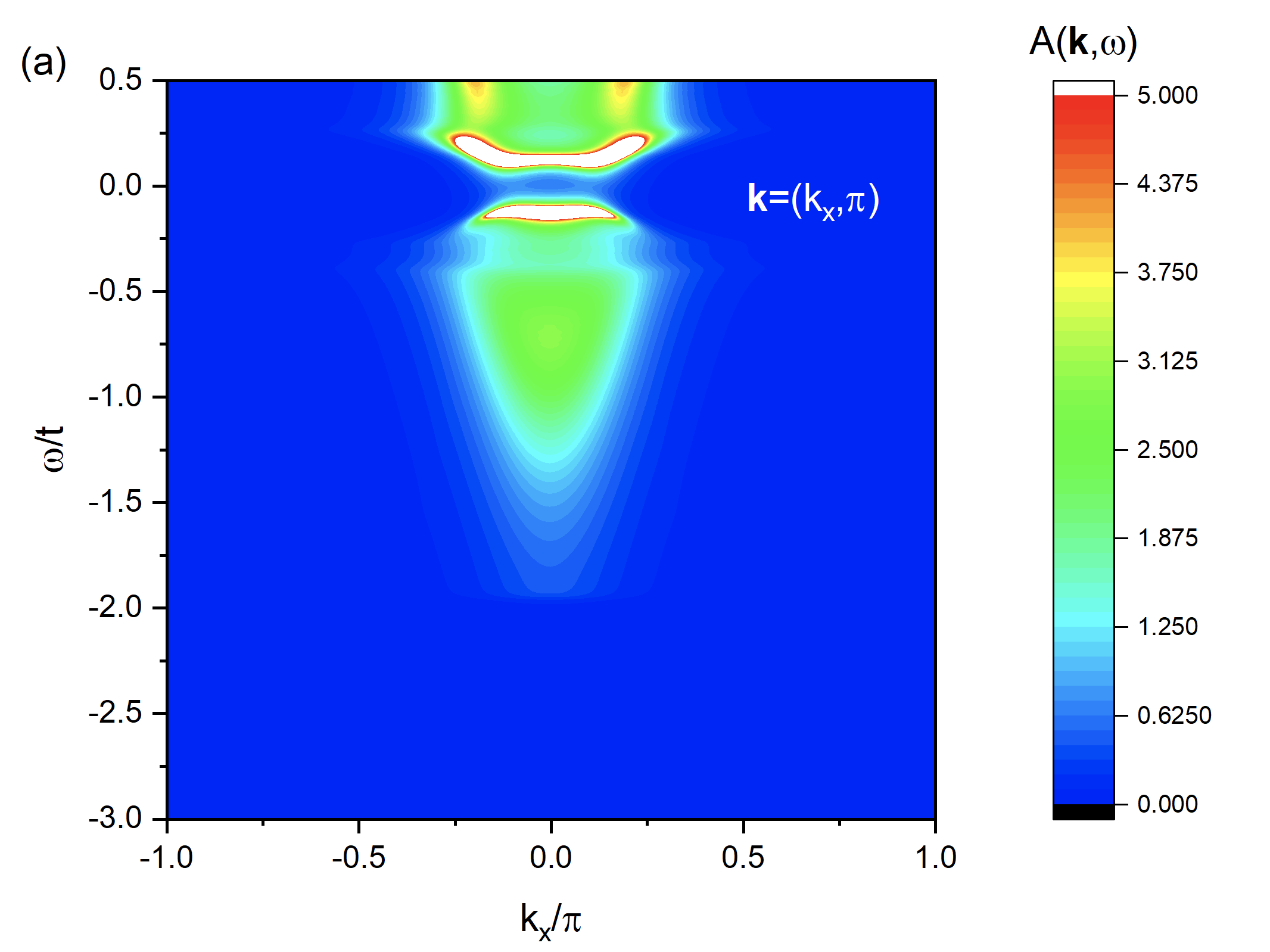}
\includegraphics[width=9cm]{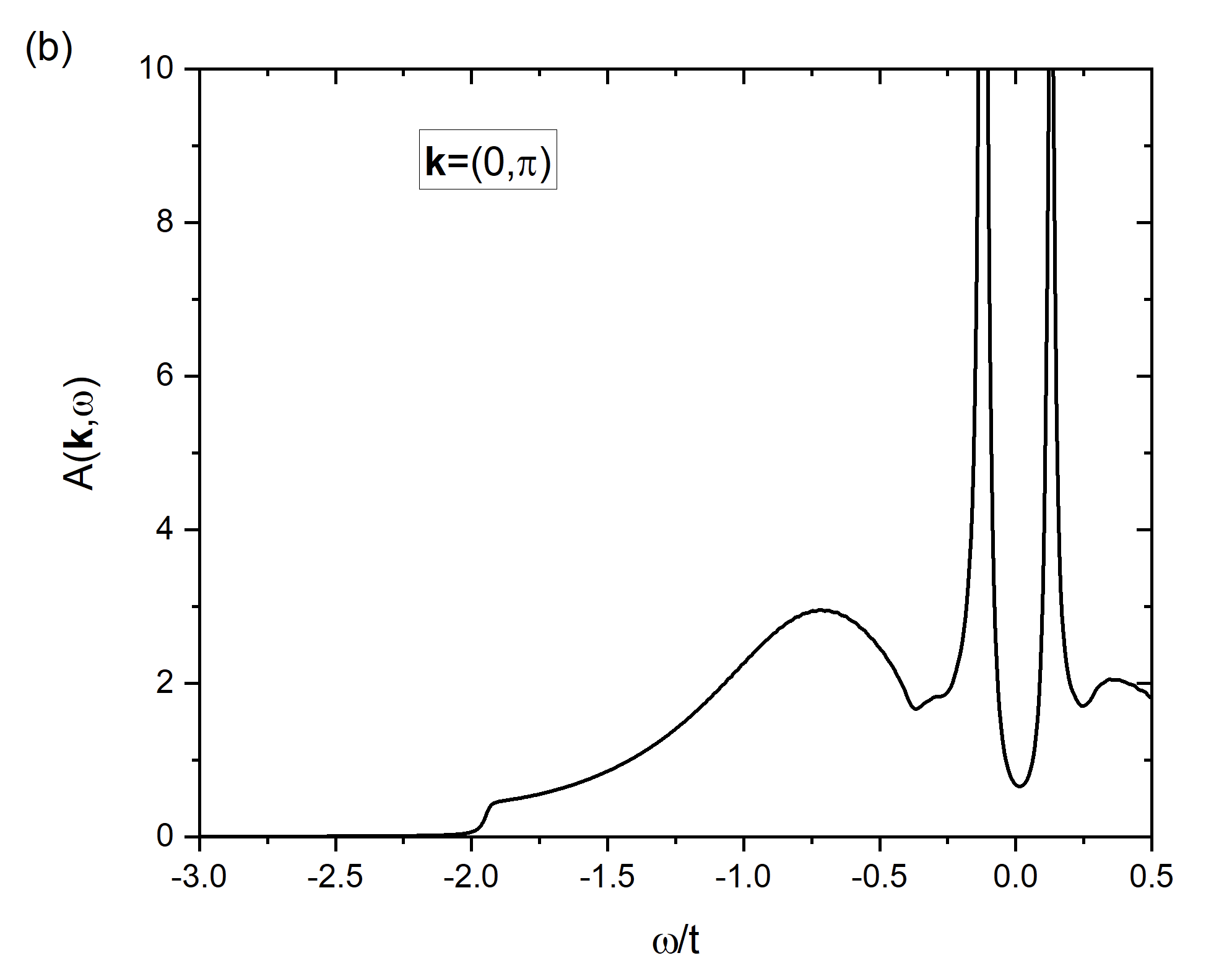}
\caption{The electron spectral function around the anti-nodal point calculated with the self-energy correction $\Sigma_{ph}$ contributed by the scattering from the $B_{1g}$ buckling mode. (a)Distribution of the electron spectral weight along  the $(-\pi,\pi)-(\pi,\pi)$ line in the superconducting state. (b) The energy distribution curve(EDC) at the anti-nodal point $\mathbf{k}=(0,\pi)$.}
\end{figure}  

The electron spectral function along the $(-\pi,\pi)-(\pi,\pi)$ line calculated with the above self-energy is shown in Fig.6. For sufficiently strong electron-phonon coupling, a clear peak-dip-hump structure also emerges in the anti-nodal region. The phenomenology is very similar to what we found in the last subsection. Most importantly, the quasiparticle peak also acquires an exceptionally flat dispersion. Accompanying this is the emergence a broad hump at higher binding energy.  By examining the expression of the electron self-energy, we find that this again can be understood as a level repulsion effect. More specifically, the broad hump is mainly contributed by scattering final state when the phonon momentum is small in magnitude and is directed along the $\mathbf{q}=(0,q_{y})$ direction. For anti-nodal quasiparticle at $\mathbf{k}=(0,\pi)$, scattering with such phonon will excite a hole in the momentum region $\mathbf{k}=(0,\pi\pm q_{y})$. It is the level repulsion with these scattering final states that induce the strong renormalization of the quasiparticle dispersion in the anti-nodal region. We note that the coupling strength with the $B_{1g}$ buckling mode reaches its maximal value at $\mathbf{q}=(0,0)$, when the scattering final state is almost degenerate with the bare quasiparticle state.

To verify this understanding, we have calculated the electron spectral function when the phonon contribution to the electron self-energy is deliberately masked if the scattering final state resides in the shaded momentum region as illustrated in Fig.7a. Indeed, the PDH structure in the resultant electron spectral function as is shown in Fig.7b  and Fig.7c is greatly suppressed. Thus, except for the fact that the dominant momentum transfer is now changed from $\mathbf{q}=(\pi,\pi)$ to $\mathbf{q}=(0,0)$ and that the momentum region involved is more extended, the mechanism for the PDH structure in the electron-phonon coupling scenario is basically the same as that in the antiferromagnetic spin fluctuation scenario.           

 \begin{figure}
\includegraphics[width=9cm]{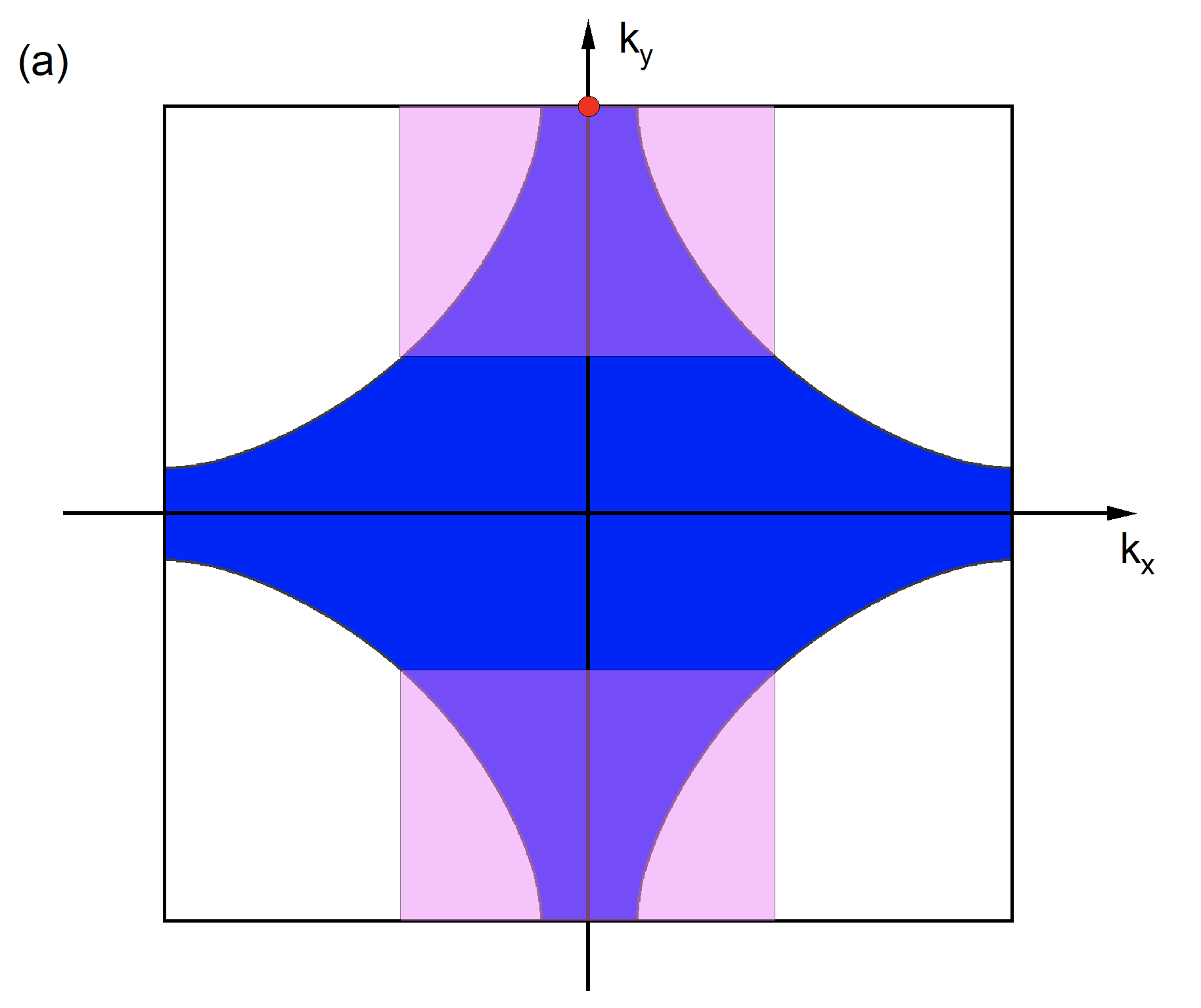}
\includegraphics[width=9cm]{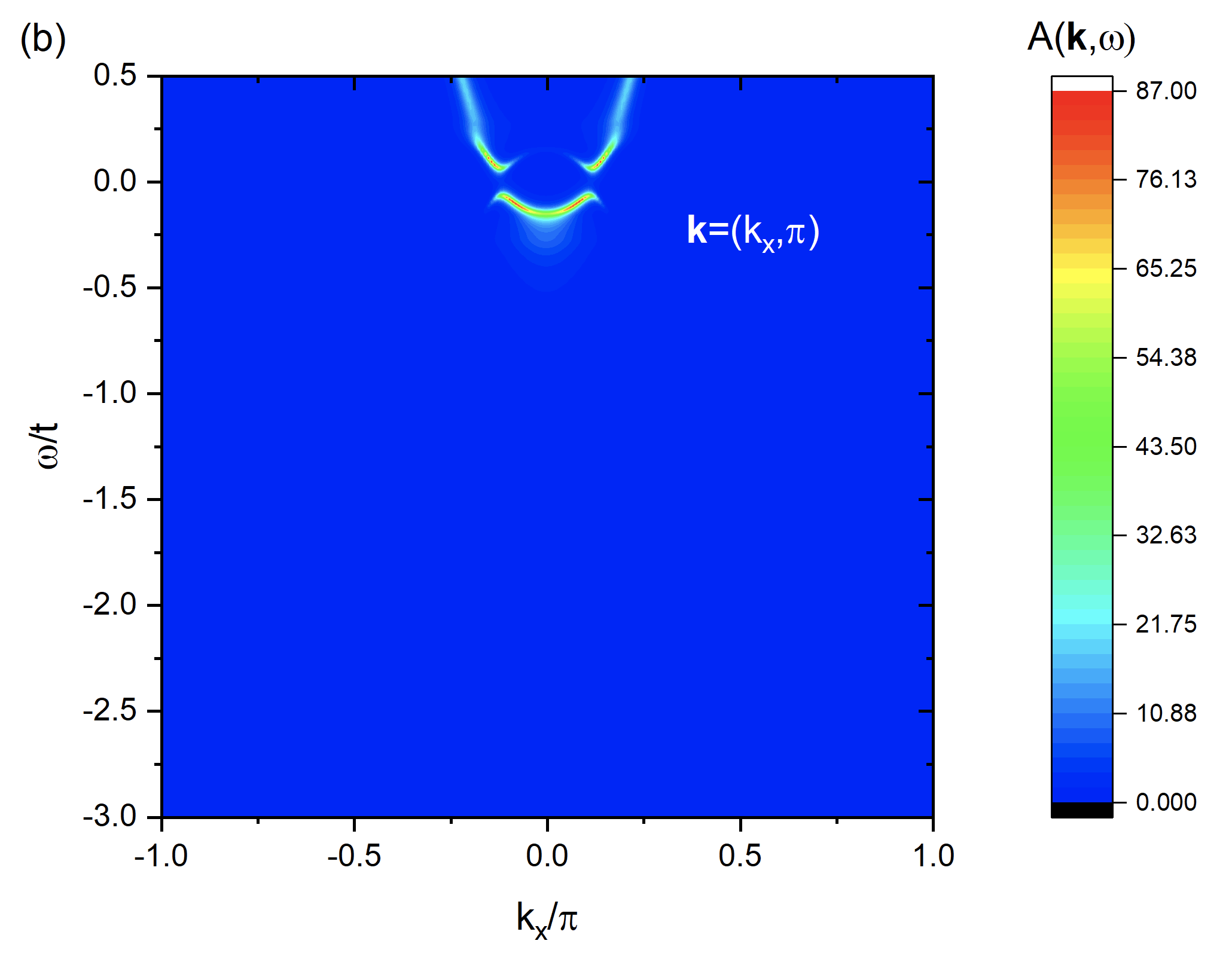}
\includegraphics[width=9cm]{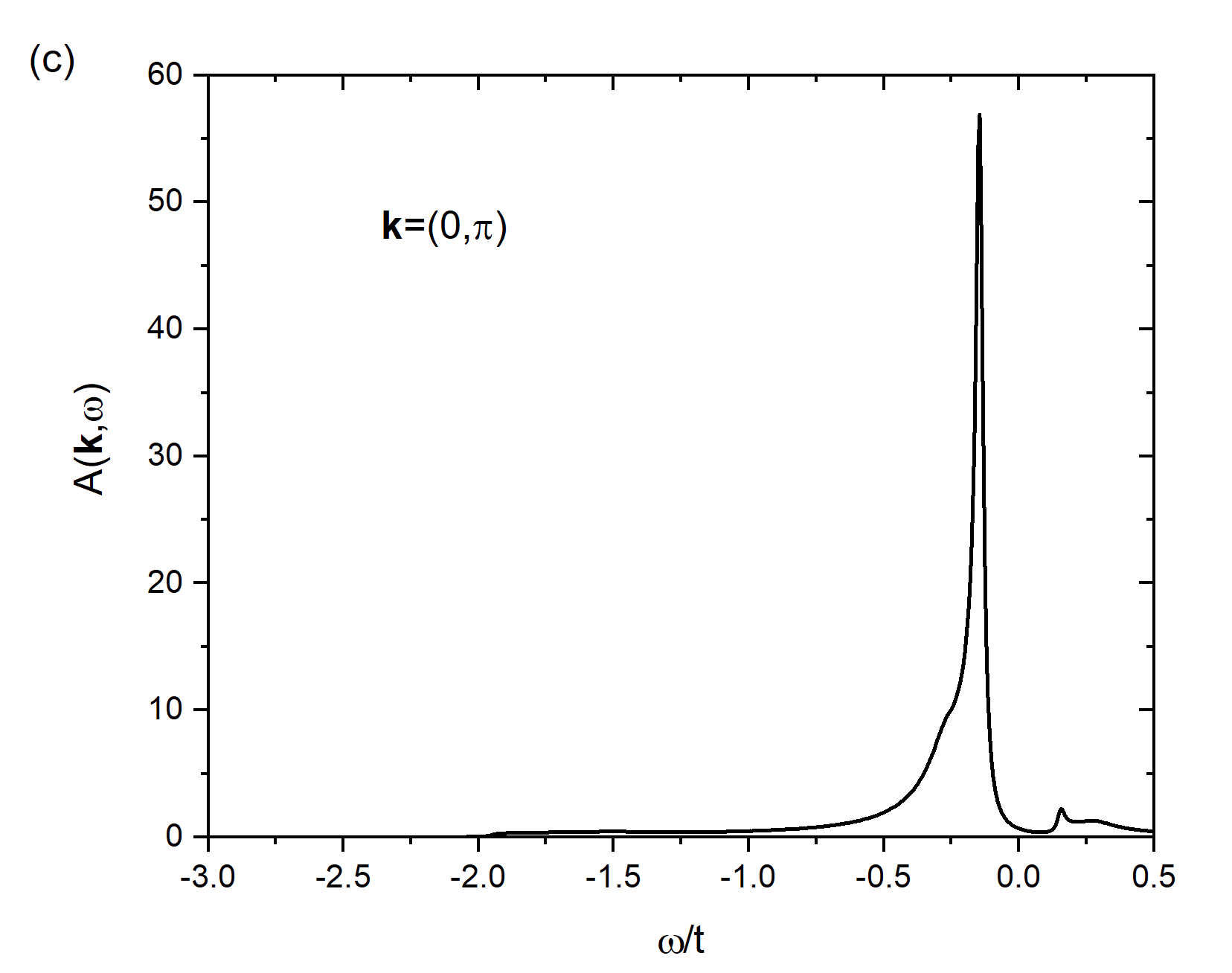}
\caption{The electron spectral function around the anti-nodal point calculated from $\Sigma_{ph}$ with the dominate momentum region of scattering final state, namely the momentum region around the same anti-nodal point in the Brillouin zone, deliberately masked. (a)The shaded area illustrated the dominate momentum region of the scattering final state in the scattering process induced by $B_{1g}$ buckling mode. (b)Distribution of the electron spectral weight along  the $(-\pi,\pi)-(\pi,\pi)$ line in the superconducting state with contribution from the shaded area masked. (c) The energy distribution curve(EDC) at the anti-nodal point $\mathbf{k}=(0,\pi)$ with contribution from the shaded area masked. The red dot denotes the position of the anti-nodal point around which the electron spectral function is calculated.}
\end{figure}

 \section{Interplay between the antiferromagnetic spin fluctuation and the electron-phonon coupling in the anti-nodal region}
 \subsection{General considerations}
 
 The discussion in the last section leads us to the following dilemma: the same PDH structure in the anti-nodal spectrum can be interpreted either in terms of the antiferromagnetic spin fluctuation scenario or the electron-phonon coupling scenario, if we treat the less well specified coupling constant $g$ or $\lambda$ as free tuning parameter. A possible clue to resolve this dilemma is to consider the doping evolution of the PDH structure, which is found to suffer a sudden suppression around the pseudogap end point with overdoping. While in the antiferromagnetic spin fluctuation scenario such a sudden suppression of the PDH structure may be linked to the dramatic change in the nature of spin fluctuation in the system as observed in RIXS measurement, much less can be said definitely within the electron-phonon coupling scenario. In fact, it is strongly debated how electron correlation effect would affect the manifestation of the electron-phonon coupling in a doped Mott insulator, or more specifically, in the pseudogap phase. Naively, the strong antiferromagnetic spin fluctuation in the pseudogap phase would compete with the electron-phonon coupling in the electron spectral weight. More specifically, the antiferromagnetic spin correlation would block electron hopping between nearest neighboring site. However, this is just the electron operator that the $B_{1g}$ buckling mode couples to. As we will demonstrate below from both mean field theory analysis and the calculation of the first order vertex correction to the electron-phonon coupling from the antiferromagnetic spin fluctuation, this is indeed the case in the $\mathbf{q}\rightarrow0$ limit. Naively, this seems to imply that the phonon contribution to the PDH structure would be suppressed in the pesudogap phase. Our calculation below shows that the answer is surprisingly the opposite. Clearly, a thorough study of the interplay between the electron-phonon coupling and the antiferromagnetic spin fluctuation in the pseudogap phase is crucial to resolve this puzzle. In the following, we will first present a mean field analysis on the interplay between electron-phonon coupling and the antiferromagnetic correlation. Here the condition $f(\mathbf{k,q})=-f(\mathbf{k+Q,q})$ of the electron-phonon coupling matrix element plays a crucial role. This will be followed a calculation of the first order vertex correction to the electron-phonon coupling from the antiferromagnetic spin fluctuation. The result strengthens again the conclusion that the electron-phonon coupling is suppressed by the antiferromagnetic spin fluctuation in the $\mathbf{q}\rightarrow 0$ limit. However, the calculation also indicates that the vertex correction from the antiferromagnetic spin fluctuation has a peculiar momentum dependence and is highly anisotropic around the $\mathbf{q}=0$ point. This feature is crucial in the analysis of the phonon contribution to the PDH structure. We then present the result of full vertex correction calculated in the ladder approximation and show that the qualitative feature of the first order calculation is preserved to the infinite order. Finally, we calculate the electron spectral function in the anti-nodal region with the intertwining effect between the electron-phonon coupling and the antiferromagnetic spin fluctuation included.

\subsection{A mean field analysis of the interplay between the electron-phonon coupling and the antiferromagnetic spin correlation}
  
Before considering the vertex correction to the electron-phonon coupling from dynamical spin fluctuation, we first consider a simplified version of the problem. For this purpose, we consider the case in which the antiferromagnetic spin fluctuation is frozen into long range antiferromagnetic order. This can be described by a MMP susceptibility with $\xi\rightarrow \infty$ and $\omega_{sf}\rightarrow 0$. In such a long range ordered background the Hamiltonian of the system takes the form\cite{Hirsch}
\begin{equation}
H_{MF}=\sum_{\mathbf{k},\sigma } \epsilon_{\mathbf{k}}c^{\dagger}_{\mathbf{k},\sigma}c_{\mathbf{k},\sigma}-\frac{gm}{2}\sum_{i,\sigma}e^{i\mathbf{Q}\cdot \mathbf{r}_{i}}\sigma c^{\dagger}_{i,\sigma}c_{i,\sigma}
\end{equation}
Here $\mathbf{Q}=(\pi,\pi)$ denotes the antiferromagnetic wave vector, $m=|\langle \vec{S}_{i}\rangle|$ is the length of the ordered moment. We now calculate the electron-phonon coupling matrix element between eigenstates of the antiferromagnetic ordered state. In the momentum space the mean field Hamiltonian of the ordered state reads
\begin{equation}
H_{MF}=\sum_{\mathbf{k}\in AFBZ,\sigma }\Psi^{\dagger}_{\mathbf{k},\sigma}\left(\begin{array}{cc}\epsilon_{\mathbf{k}} & -\frac{gm}{2}\sigma \\-\frac{gm}{2}\sigma & \epsilon_{\mathbf{k+Q}}\end{array}\right) \Psi_{\mathbf{k},\sigma}
\end{equation}
in which $AFBZ$ denotes the Brillouin zone of the antiferromagnetic ordered state,
\begin{equation}
\Psi_{\mathbf{k},\sigma}=\left(\begin{array}{c}c_{\mathbf{k},\sigma} \\c_{\mathbf{k+Q},\sigma}\end{array}\right)
\end{equation}
Diagonalizing the Hamiltonian with the following unitary transformation
\begin{equation}
\Psi_{\mathbf{k},\sigma}=U_{\mathbf{k},\sigma}\bm{\gamma}_{\mathbf{k},\sigma}=\left(\begin{array}{cc}u_{\mathbf{k},\sigma} & -v_{\mathbf{k},\sigma} \\v_{\mathbf{k},\sigma} & u_{\mathbf{k},\sigma}\end{array}\right)\left(\begin{array}{c}\gamma_{\mathbf{k,+},\sigma} \\\gamma_{\mathbf{k,-},\sigma}\end{array}\right)\nonumber
\end{equation}
we are led to
\begin{equation}
H_{MF}=\sum_{\mathbf{k}\in AFBZ,s=\pm,\sigma}\tilde{\epsilon}_{\mathbf{k},s}\gamma^{\dagger}_{\mathbf{k},s,\sigma}\gamma_{\mathbf{k},s,\sigma}
\end{equation}
in which  
\begin{equation}
\tilde{\epsilon}_{\mathbf{k},\pm}=\frac{( \epsilon_{\mathbf{k}}+ \epsilon_{\mathbf{k+Q}})\pm\sqrt{( \epsilon_{\mathbf{k}}- \epsilon_{\mathbf{k+Q}})^{2}+(gm)^{2}/4}}{2}
\end{equation}
is the quasiparticle energy in the antiferromagnetic ordered background.

The electron-phonon coupling Hamiltonian can be rewritten in terms of the $\bm{\gamma}$ operators as follows
\begin{eqnarray}
H_{SSH}&=&\frac{1}{\sqrt{N}}\sum_{\mathbf{k,q},\sigma}f(\mathbf{k,q})u^{B}_{\mathbf{q}}c^{\dagger}_{\mathbf{k+q},\sigma}c_{\mathbf{k},\sigma}\nonumber\\
&=&\frac{1}{\sqrt{N}}\sum_{\mathbf{k}\in AFBZ,\mathbf{q},\sigma}f(\mathbf{k,q})u^{B}_{\mathbf{q}}\Psi^{\dagger}_{\mathbf{k+q},\sigma}\tau_{3}\Psi_{\mathbf{k},\sigma}\nonumber\\
\end{eqnarray}
in which 
\begin{equation}
\tau_{3}=\left(\begin{array}{cc}1 & 0 \\0 & -1\end{array}\right)
\end{equation}
In the derivation we have used the identity $f(\mathbf{k,q})=-f(\mathbf{k+Q,q})$ and extended the definition of $\Psi_{\mathbf{k},\sigma}$ from the $AFBZ$ to the full Brillouin zone of the square lattice in such a way that
\begin{equation}
\Psi_{\mathbf{k+Q},\sigma}=\tau_{1}\Psi_{\mathbf{k},\sigma}
\end{equation}
in which 
\begin{equation}
\tau_{1}=\left(\begin{array}{cc}0 & 1 \\1 & 0\end{array}\right)
\end{equation}
Inserting the unitary transformation $\Psi_{\mathbf{k},\sigma}=U_{\mathbf{k},\sigma}\bm{\gamma}_{\mathbf{k},\sigma}$ we have
\begin{eqnarray}
H_{SSH}&=&\frac{1}{\sqrt{N}}\sum_{\mathbf{k}\in AFBZ,\mathbf{q},\sigma}f(\mathbf{k,q})u^{B}_{\mathbf{q}}\bm{\gamma}^{\dagger}_{\mathbf{k+q},\sigma}\mathbf{M_{k,q,\sigma}}\bm{\gamma}_{\mathbf{k},\sigma}\nonumber\\
\end{eqnarray}
in which
\begin{equation}
\mathbf{M_{k,q,\sigma}}=U^{\dagger}_{\mathbf{k+q},\sigma}\tau_{3}U_{\mathbf{k},\sigma}
\end{equation}
Here the definition of $\bm{\gamma}_{\mathbf{k},\sigma}$ and $U_{\mathbf{k},\sigma}$ has also been extended to the full Brillouin zone of the square lattice in such a way that
\begin{eqnarray}
\bm{\gamma}_{\mathbf{k+Q},\sigma}&=&\bm{\gamma}_{\mathbf{k},\sigma}\nonumber\\
U_{\mathbf{k+Q},\sigma}&=&\tau_{1}U_{\mathbf{k},\sigma}
\end{eqnarray}
The matrix elements of $\mathbf{M_{k,q,\sigma}}$ are given as follows. For $\mathbf{k,k+q}\in AFBZ$, we have
\begin{equation}
\mathbf{M_{k,q,\sigma}}=\left(\begin{array}{cc}M_{11} & -M_{12} \\-M_{12} & -M_{11}\end{array}\right)
\end{equation}
in which 
\begin{eqnarray}
M_{11}&=&u_{\mathbf{k+q,\sigma}}u_{\mathbf{k,\sigma}}-v_{\mathbf{k+q,\sigma}}v_{\mathbf{k,\sigma}}\nonumber\\
M_{12}&=&u_{\mathbf{k+q,\sigma}}v_{\mathbf{k,\sigma}}+v_{\mathbf{k+q,\sigma}}u_{\mathbf{k,\sigma}}
\end{eqnarray}
While for $\mathbf{k}\in AFBZ$ but $\mathbf{k+q}\notin AFBZ$, we have
\begin{equation}
\mathbf{M_{k,q,\sigma}}=\left(\begin{array}{cc}M_{11} & -M_{12} \\M_{12} & M_{11}\end{array}\right)
\end{equation}
in which 
\begin{eqnarray}
M_{11}&=&v_{\mathbf{k+q+Q,\sigma}}u_{\mathbf{k,\sigma}}-u_{\mathbf{k+q+Q,\sigma}}v_{\mathbf{k,\sigma}}\nonumber\\
M_{12}&=&u_{\mathbf{k+q+Q,\sigma}}u_{\mathbf{k,\sigma}}+v_{\mathbf{k+q+Q,\sigma}}v_{\mathbf{k,\sigma}}
\end{eqnarray}
Note that if $\mathbf{k+q}\notin AFBZ$, then $\mathbf{k+q+Q}\in AFBZ$.  

Let us focus first on the intra-band matrix element of the electron-phonon coupling and consider the $\mathbf{q}=0$ limit.  Clearly,  the intra-band matrix element of the electron-phonon coupling as given by $f(\mathbf{k,q})M_{11}$ is suppressed on the whole fermi surface by the antiferromagnetic order. In particular, it becomes exactly zero at the hot spots where $\epsilon_{\mathbf{k}}=\epsilon_{\mathbf{k+Q}}=0$. Clearly, the suppression in the intra-band matrix element is compensated by the increase in the inter-band matrix element of the electron-phonon coupling, which is given by $f(\mathbf{k,q})M_{12}$. Indeed, it is easy to verify that 
\begin{equation}
M^{2}_{11}+M^{2}_{12}=1
\end{equation}
Such a suppression in the intra-band matrix element of the electron-phonon coupling is easily understood from the real space perspective, since the nearest neighboring hopping of the electron that the $B_{1g}$ buckling mode couples to is suppressed by the antiferromagnetic order.

The intra-band and the inter-band electron-phonon scattering dominate different physical processes. More specifically, the intra-band electron-phonon scattering is expected to dominate the phonon contribution to the DC transport scattering rate, while the inter-band electron-phonon scattering is expected to be important in the optical absorption process. According to the discussion above, we expect the phonon contribution to the DC transport scattering rate to be suppressed in an antiferromagnetic correlated background. This may explain the violation of the Mathiessen's rule in the strange metal phase of cuprate superconductors, namely the absence of any discernible change in the slope of the linear-in-T resistivity at a temperature scale when we expect the electron-phonon coupling to play an essential role\cite{Liu1}. On the other hand, we expect that the phonon side band in the optical absorption spectrum to be enhanced with the strengthen of the antiferromagnetic correlation in the system.   

 \subsection{The first order vertex correction to the electron-phonon coupling from the antiferromagnetic spin fluctuation}
Now we consider the vertex correction to the electron-phonon coupling from dynamical spin fluctuation. To simplify the discussion, here we will stick to the normal state. The Feynman diagram of the first order vertex correction to the electron-phonon coupling is illustrated in Fig.8. Following the Feynman's rule, the first order vertex correction reads
 \begin{eqnarray}
\Gamma^{(1)}(\mathbf{k},i\nu_{n},\mathbf{q},i\omega_{n})&=&-\frac{3g^{2}}{4N\beta}\sum_{\mathbf{q'},i\omega_{n'}}f(\mathbf{k-q'},\mathbf{q})\nonumber\\
&\times&\chi(\mathbf{q'},i\omega_{n'})g(\mathbf{k-q'},i\nu_{n}-i\omega_{n'})\nonumber\\
&\times&g(\mathbf{k+q-q'},i\nu_{n}+i\omega_{n}-i\omega_{n'})\nonumber\\
\end{eqnarray}
in which 
\begin{equation}
g(\mathbf{k},i\nu_{n})=\frac{1}{i\nu_{n}-\epsilon_{\mathbf{k}}}
\end{equation}
denotes the electron Green's function in the normal state. Inserting the spectral representation of $\chi(\mathbf{q},i\omega_{n})$ and completing the summation over the Matsubara frequency we obtain
\begin{eqnarray}
&&\Gamma^{(1)}(\mathbf{k},i\nu_{n},\mathbf{q},i\omega_{n})=\frac{3g^{2}}{8\pi N}\sum_{\mathbf{q'}}f(\mathbf{k-q'},\mathbf{q})\nonumber\\
&\times&\int_{0}^{\infty}d \omega' R(\mathbf{q'},\omega') K(\mathbf{k,q,q'},i\nu_{n},i\omega_{n},\omega')
\end{eqnarray} 
with
\begin{eqnarray}
K(\mathbf{k,q,q'},i\nu_{n},i\omega_{n},\omega')&=&g_{4}g_{5}+n_{B}(\omega')(g_{1}g_{2}+g_{4}g_{5})\nonumber\\
&+&(1-f(\epsilon_{\mathbf{k-q'}}))(g_{1}g_{3}-g_{3}g_{4})\nonumber\\
&+&(1-f(\epsilon_{\mathbf{k+q-q'}}))(g_{5}g_{3}-g_{3}g_{2})\nonumber\\
\end{eqnarray}    
in which
\begin{eqnarray}
g_{1}&=&\frac{1}{i\nu_{n}-\omega'-\epsilon_{\mathbf{k-q'}}}\nonumber\\
g_{2}&=&\frac{1}{i\nu_{n}+i\omega_{n}-\omega'-\epsilon_{\mathbf{k+q-q'}}}\nonumber\\
g_{3}&=&\frac{1}{i\omega_{n}+\epsilon_{\mathbf{k-q'}}-\epsilon_{\mathbf{k+q-q'}}}\nonumber\\
g_{4}&=&\frac{1}{i\nu_{n}+\omega'-\epsilon_{\mathbf{k-q'}}}\nonumber\\
g_{5}&=&\frac{1}{i\nu_{n}+i\omega_{n}+\omega'-\epsilon_{\mathbf{k+q-q'}}}\nonumber\\
\end{eqnarray}

Before presenting the full numerical result, let us first analyze the behavior of the vertex correction in the $\mathbf{q}=0$ limit. To further simplify the analysis, we take the $\omega_{n}=0$ and $\nu_{n}\rightarrow0$ limit, in which case the vertex correction reduces to
\begin{eqnarray}
&&\lim_{\nu_{n}\rightarrow 0}\Gamma^{(1)}(\mathbf{k},i\nu_{n},\mathbf{q}=0,i\omega_{n}=0)\nonumber\\
&=&\frac{3g^{2}}{8N\pi}\sum_{\mathbf{q'}}f(\mathbf{k-q'},\mathbf{q}=0)\nonumber\\
&\times&\int_{0}^{\infty}d \omega' R(\mathbf{q'},\omega')\{ \frac{1-f(\epsilon_{\mathbf{k-q'}})}{(\omega'+\epsilon_{\mathbf{k-q'}})^2}+\frac{f(\epsilon_{\mathbf{k-q'}})}{(\omega'-\epsilon_{\mathbf{k-q'}})^2}\} \nonumber\\
\end{eqnarray} 
Since the antiferromagnetic spin fluctuation is dominated by spectral weight around $\mathbf{Q}=(\pi,\pi)$ and that $f(\mathbf{k,q})=-f(\mathbf{k+Q,q})$, we see that the vertex correction from the antiferromagnetic spin fluctuation always has a sign that is opposite to the bare electron-phonon coupling vertex $f(\mathbf{k,q})$. Although such an analysis is made in the $\omega_{n}=0$ and $\nu_{n}\rightarrow0$ limit, the numerical calculation indicates that first order vertex correction from the antiferromagnetic spin fluctuation is strongly negative for small $\omega_{n}$ and $\nu_{n}$ when $\mathbf{q}=0$(see Fig.9). Thus, the electron-phonon coupling strength in the $\mathbf{q}=0$ limit is always suppressed by the vertex correction form the antiferromagnetic spin fluctuation at low energy. Naively, this seems to imply that the phonon contribution to the PDH structure would be suppressed in the pseudogap phase in which antiferromagnetic fluctuation of the local moment is strongly enhanced, since the anti-nodal quasiparticle couples mainly to the long wavelength $B_{1g}$ phonon. If this naive argument is indeed relevant, then the sudden suppression of the PDH structure around the pseudogap end point upon overdoping would be a strong evidence against the phonon scenario of the PDH structure. The following discussion indicates that the answer to this subtle problem is surprisingly the opposite.
\begin{figure}
\includegraphics[width=8cm]{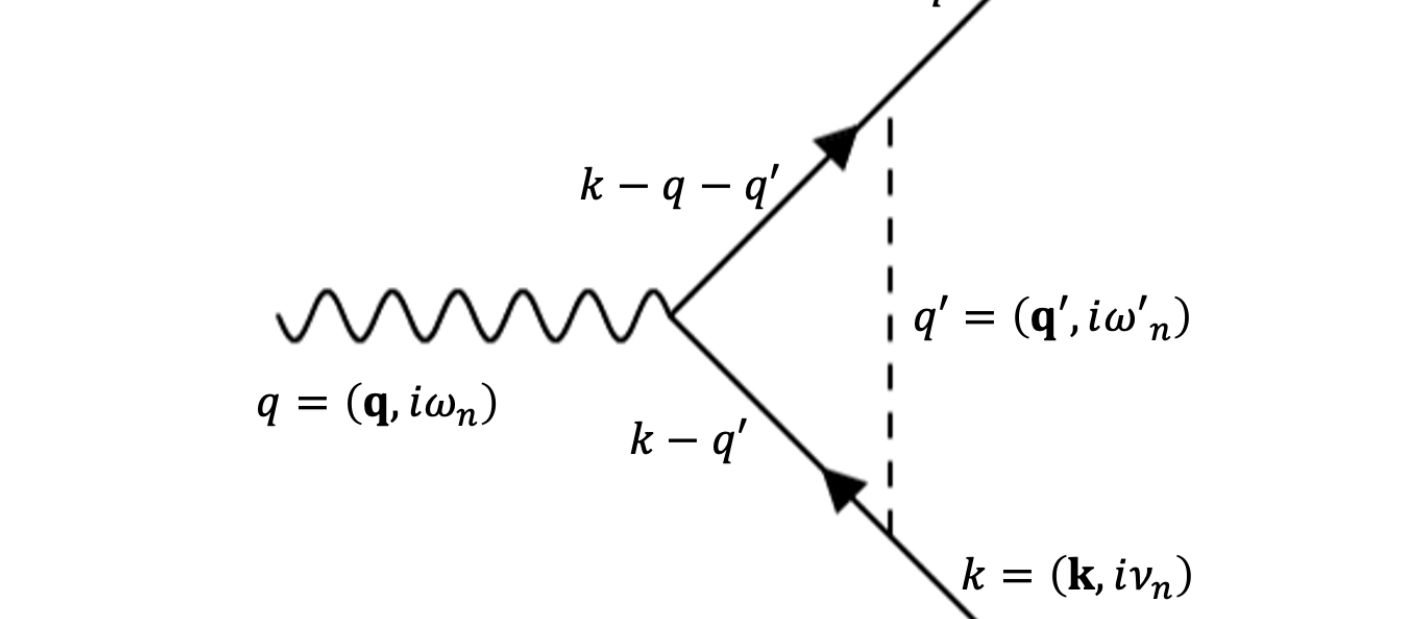}
\caption{The lowest order vertex correction to the electron-phonon coupling from the antiferromagnetic spin fluctuation. Here the solid lines denote the electron Green's function, the dashed line denotes the propagator of the antiferromagnetic spin fluctuation, and the wavy line denotes the phonon propagator.}
\end{figure}  

 \begin{figure}
\includegraphics[width=8cm]{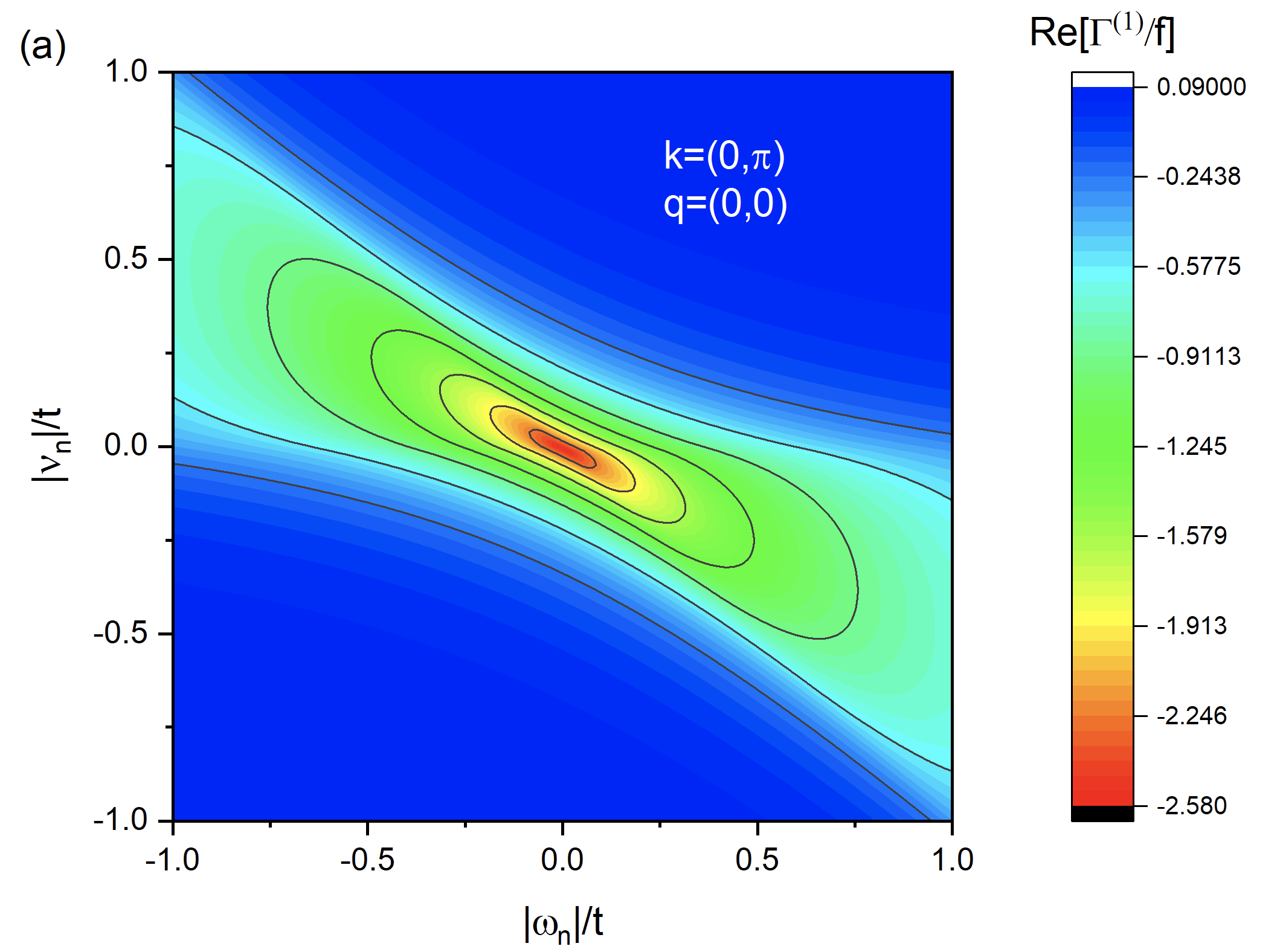}
\caption{The frequency dependence of the first order vertex correction to the electron-phonon coupling from the antiferromagnetic spin fluctuation at $\mathbf{q}=0$. The strongly negative vertex correction at small $\omega_{n}$ and $\nu_{n}$ can be attributed to the coalescence of the pole of the two electron Green's function $g(\mathbf{k-q'},i\nu_{n}-i\omega_{n'})$ and $g(\mathbf{k+q-q'},i\nu_{n}+i\omega_{n}-i\omega_{n'})$ in Eq.51 in this limit.}
\end{figure}  

\subsection{The nontrivial momentum dependence of the vertex correction}
The key problem with the above naive argument is that the vertex correction from the antiferromagnetic spin fluctuation actually has a highly non-trivial momentum dependence around $\mathbf{q}=0$. In fact, the strongly negative vertex correction at $\mathbf{q}=0$ discussed in the last subsection can be attributed to the coalescence of poles of the two electron Green's function $g(\mathbf{k-q'},i\nu_{n}-i\omega_{n'})$ and $g(\mathbf{k+q-q'},i\nu_{n}+i\omega_{n}-i\omega_{n'})$ appearing in Eq.51. The first order vertex correction is thus in a sense singular at $\mathbf{q}=0$. Such singular behavior will be eventually removed by finite quasiparticle broadening effect. Away from $\mathbf{q}=0$, $\Gamma^{(1)}(\mathbf{k},i\nu_{n},\mathbf{q}=0,i\omega_{n})$ is found to be highly anisotropic. Such an anisotropic momentum dependence originates from the highly anisotropic quasiparticle dispersion around the anti-nodal point, which is a saddle point of the quasiparticle dispersion.

The integration over the spin fluctuation frequency $\omega'$ and the summation over the spin fluctuation momentum $\mathbf{q'}$ in Eq. 53 can be readily done to obtain the numerical result of first order vertex correction $\Gamma^{(1)}(\mathbf{k},i\nu_{n},\mathbf{q},i\omega_{n})$. The result is presented in Fig.10. As can be seen from the figure, the suppression of the electron-phonon vertex is limited to a small and very anisotropic region around $\mathbf{q}=0$. In the more extended momentum region, the vertex correction from the antiferromagnetic spin fluctuation is seen to strongly peak in two crescent-shape regions centered at $\mathbf{q}=(0,\pm q_{y,0})$. The crescent shape of these two regions can be attributed to the quasiparticle dispersion of the scattering final state of the anti-nodal quasiparticle by the antiferromagnetic spin fluctuation, namely, $\epsilon_\mathbf{k+Q}$.

 \begin{figure}
\includegraphics[width=8cm]{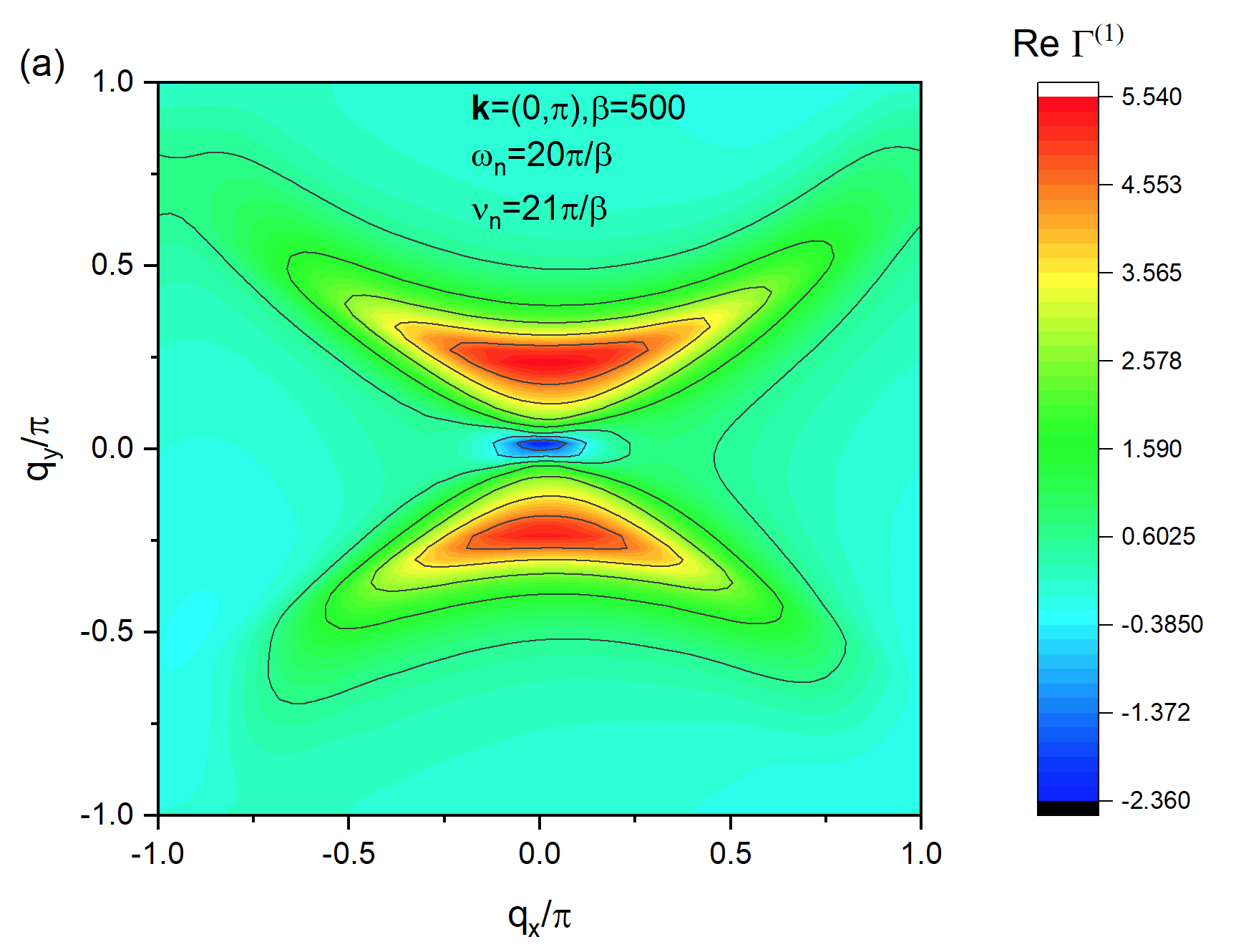}
\includegraphics[width=8cm]{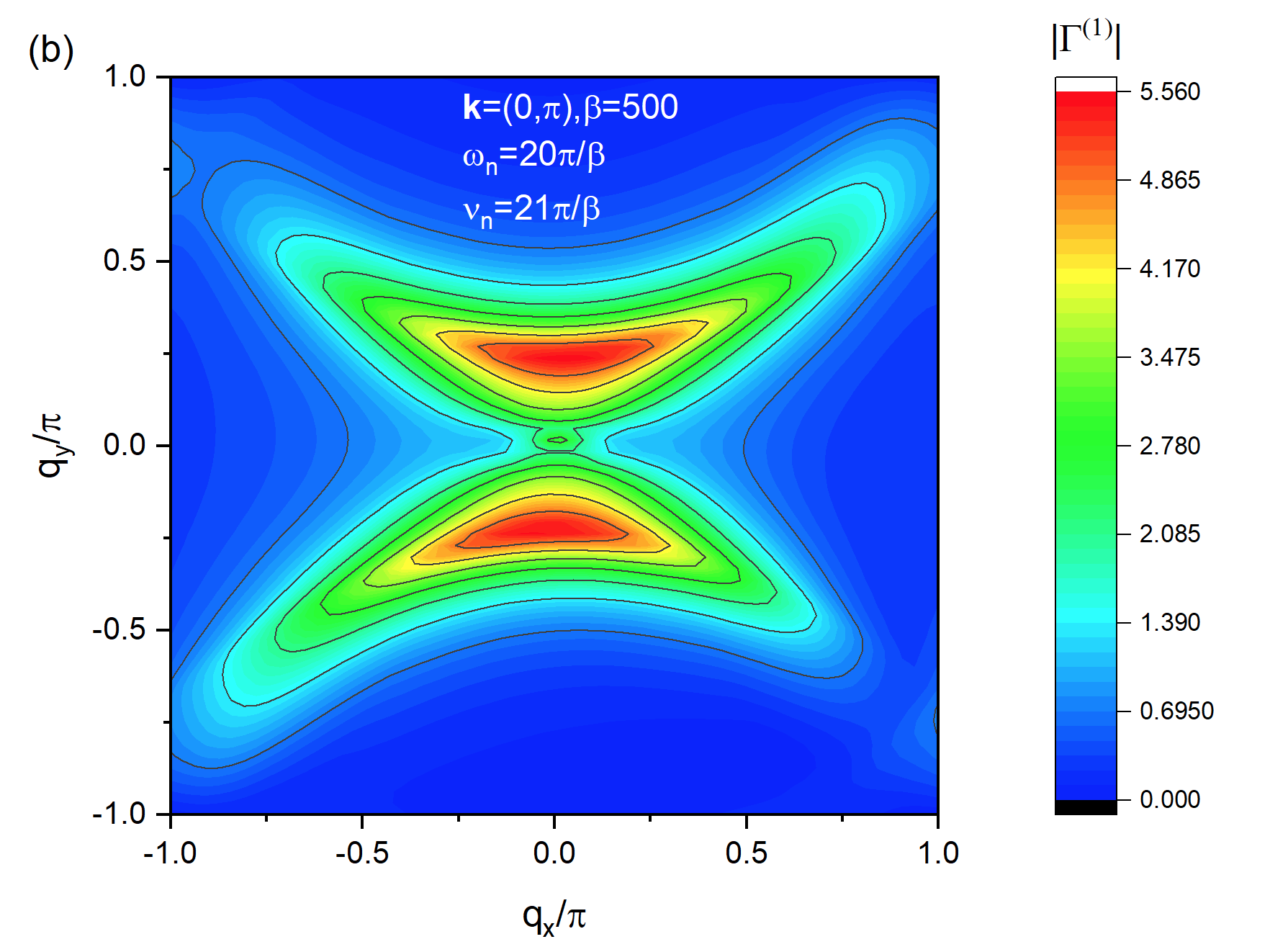}
\caption{The momentum dependence of the first order vertex correction to the electron-phonon coupling from the antiferromagnetic spin fluctuation. (a)The real part of the vertex correction, (b)The absolute value of the vertex correction. Note that the singular behavior around $\mathbf{q}=0$ is related to the coalescence of the Green's function poles and will be eventually smoothed out by finite quasiparticle life time broadening.}
\end{figure}  

Since the crescent regions are just the momentum region that dominate the phonon contribution to the high energy hump structure, it is reasonable to expect that the vertex correction from the antiferromagnetic spin fluctuation will enhance the phonon contribution to the PDH structure. This is totally opposite to the expectation from the naive argument presented in the last two subsections. However, before leaving this section and presenting the electron spectral function with the vertex correction effect included, we will first present a ladder approximation for the full vertex correction function. The purpose of this calculation is to see if the characteristic we find in the first order vertex function can be preserved at infinite perturbative order. 

 \subsection{Summation of ladder diagram and the non-crossing approximation to the full vertex function}

The self-consistent equation for the full vertex function under the ladder approximation is illustrated in Fig.11. Following the Feynman's rule, the analytical expression for the full vertex function is given by
 \begin{eqnarray}
 &&\Gamma(\mathbf{k},i\nu_{n},\mathbf{q},i\omega_{n})=f(\mathbf{k,q})\nonumber\\
 &-&\frac{3g^{2}}{4N\beta}\sum_{\mathbf{q'},i\omega'_{n}}\Gamma(\mathbf{k-q'},i\nu_{n}-i\omega'_{n},\mathbf{q},i\omega_{n})\chi(\mathbf{q'},i\omega'_{n})\nonumber\\
 &\times&g(\mathbf{k-q'},i\nu_{n}-i\omega'_{n})g(\mathbf{k-q'+q},i\nu_{n}-i\omega'_{n}+i\omega_{n})\nonumber\\
 \end{eqnarray}
 This equation can be solved iteratively by numerical summation over the Matsubara frequency. Since both the electron propagator $g(\mathbf{k},i\nu_{n})$ and the spin fluctuation propagator $\chi(\mathbf{q},i\omega_{n})$ vanish in the high frequency limit, we expect that 
 \begin{equation}
 \tilde{\Gamma}(\mathbf{k},i\nu_{n},\mathbf{q},i\omega_{n})= \Gamma(\mathbf{k},i\nu_{n},\mathbf{q},i\omega_{n})-f(\mathbf{k,q}),
 \end{equation}
namely the difference between the full vertex function and the bare electron-phonon vertex to decay rapidly with the electron frequency $|\nu_{n}|$. In the following, we will focus on $\tilde{\Gamma}(\mathbf{k},i\nu_{n},\mathbf{q},i\omega_{n})$ and assume it to decay sufficiently fast with $|\nu_{n}|$. It is easy to see that $\tilde{\Gamma}(\mathbf{k},i\nu_{n},\mathbf{q},i\omega_{n})$ satisfies the following equation
 \begin{eqnarray}
&&\tilde{\Gamma}(\mathbf{k},i\nu_{n},\mathbf{q},i\omega_{n})=\Gamma^{(1)}(\mathbf{k},i\nu_{n},\mathbf{q},i\omega_{n})\nonumber\\
&-&\frac{3g^{2}}{4N\beta}\sum_{\mathbf{q'},i\omega'_{n}}\tilde{\Gamma}(\mathbf{k-q'},i\nu_{n}-i\omega'_{n},\mathbf{q},i\omega_{n})\chi(\mathbf{q'},i\omega'_{n})\nonumber\\
 &\times&g(\mathbf{k-q'},i\nu_{n}-i\omega'_{n})g(\mathbf{k-q'+q},i\nu_{n}-i\omega'_{n}+i\omega_{n})\nonumber\\
 \end{eqnarray}
 At fixed $\mathbf{q}$ and $i\omega_{n}$, we can iterate the above equation on a finite momentum mesh to generate a self-consistent solution for the $\mathbf{k}$ and $i\nu_{n}$ dependence of $\tilde{\Gamma}(\mathbf{k},i\nu_{n},\mathbf{q},i\omega_{n})$. 
 
  \begin{figure}
\includegraphics[width=8cm]{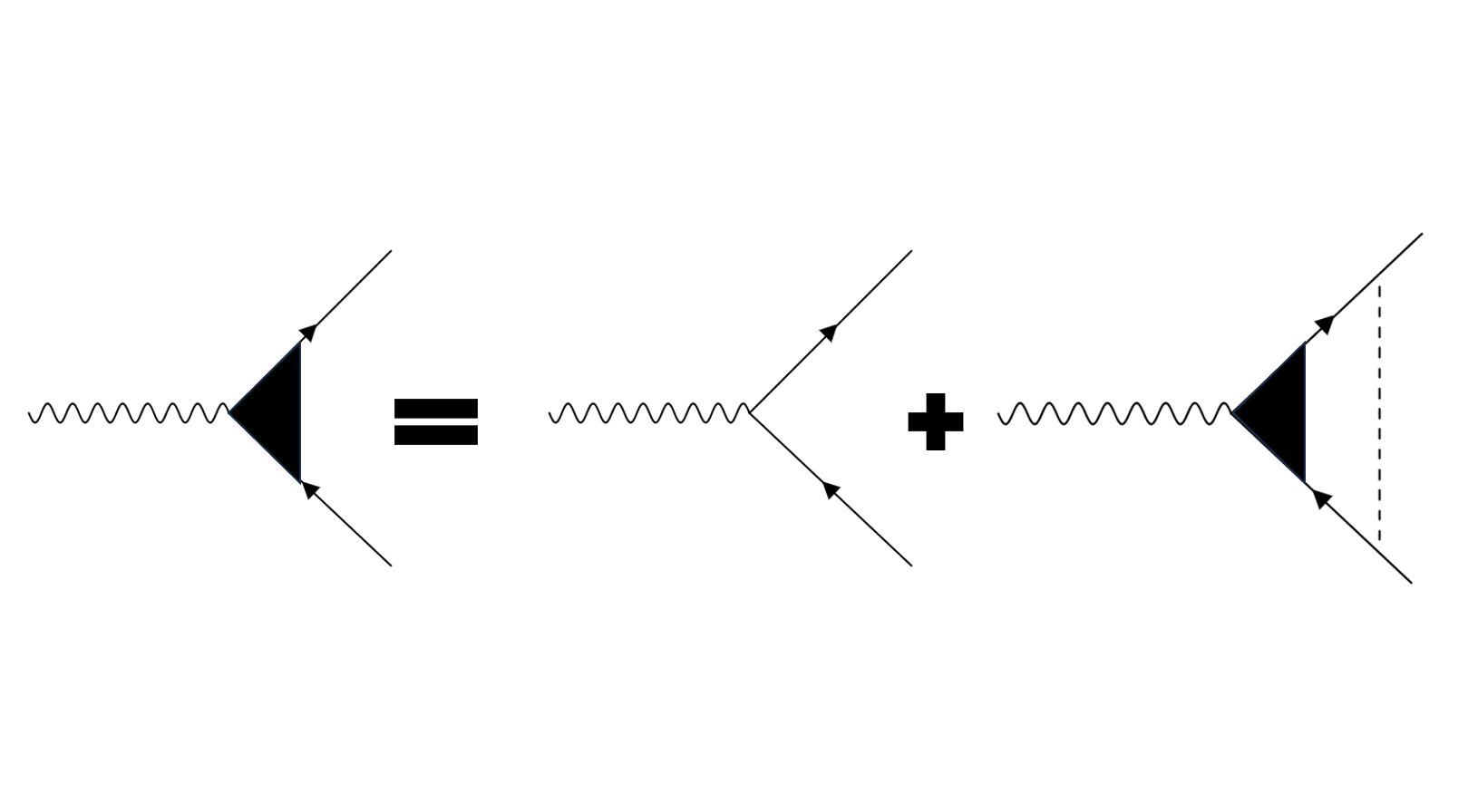}
\caption{Feynman diagram for the full electron-phonon vertex function under the ladder approximation for the scattering caused by the antiferromagnetic spin fluctuation. Here the solid lines denote the electron Green's function, the dashed line denotes the propagator of the antiferromagnetic spin fluctuation, the wavy lines denote the phonon propagator, and the shaded triangles denote the full electron-phonon vertex function.}
\end{figure}  
 
Our numerical calculation of the full vertex function is carried out on a $24\times24$ lattice with periodic boundary condition. We have truncated the $i\nu_{n}$ dependence of $\tilde{\Gamma}(\mathbf{k},i\nu_{n},\mathbf{q},i\omega_{n})$ at a cutoff frequency of $|\nu_{n}|=2\pi t$, which is found to be sufficiently large. The dependence of $\tilde{\Gamma}(\mathbf{k},i\nu_{n},\mathbf{q},i\omega_{n})$ with $\mathbf{q}$ is shown in Fig.12 for $\mathbf{k}=(0,\pi)$. Clearly, the non-trivial momentum dependence of the first order vertex function is preserved in the full vertex function.  
 
\begin{figure}
\includegraphics[width=8cm]{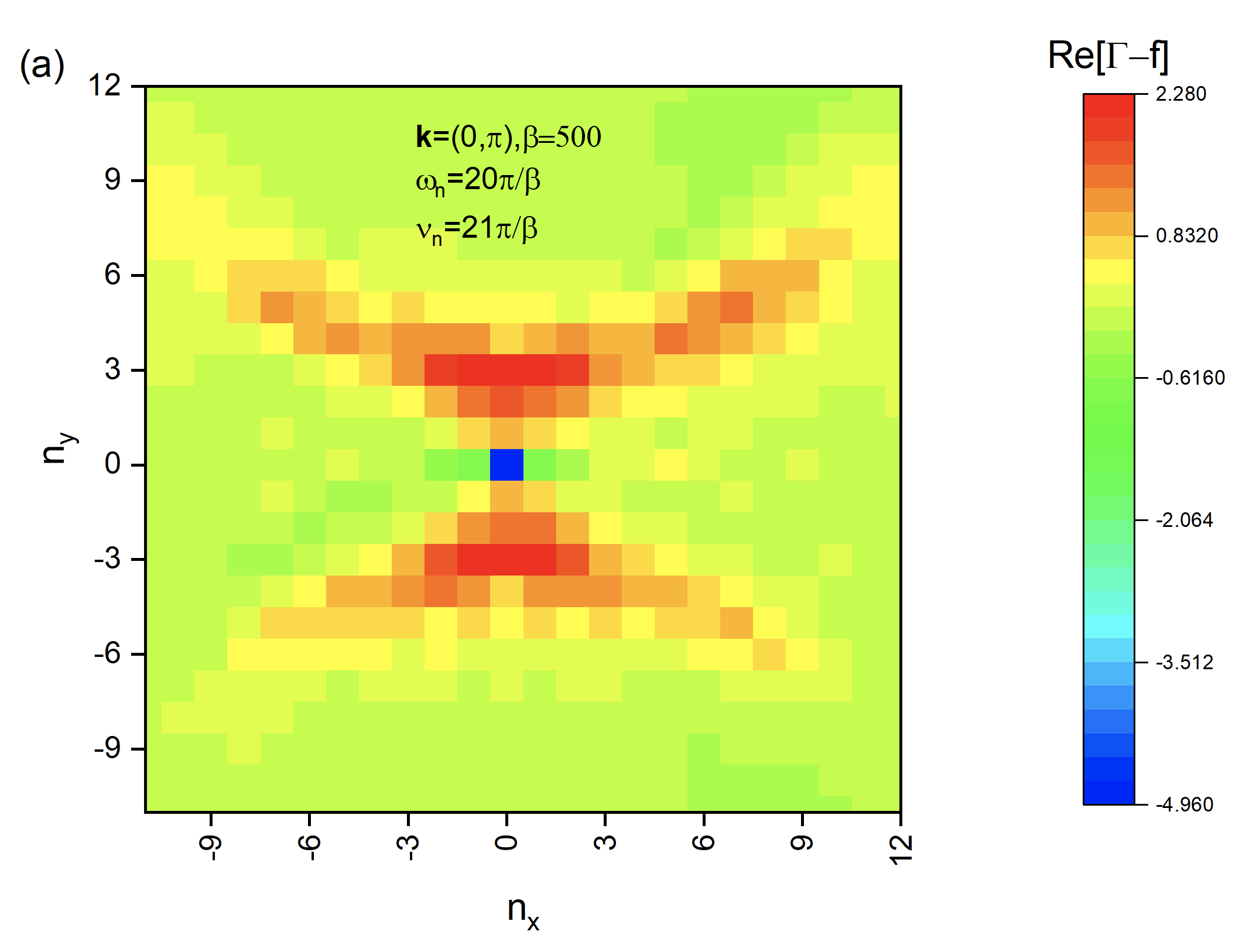}
\includegraphics[width=8cm]{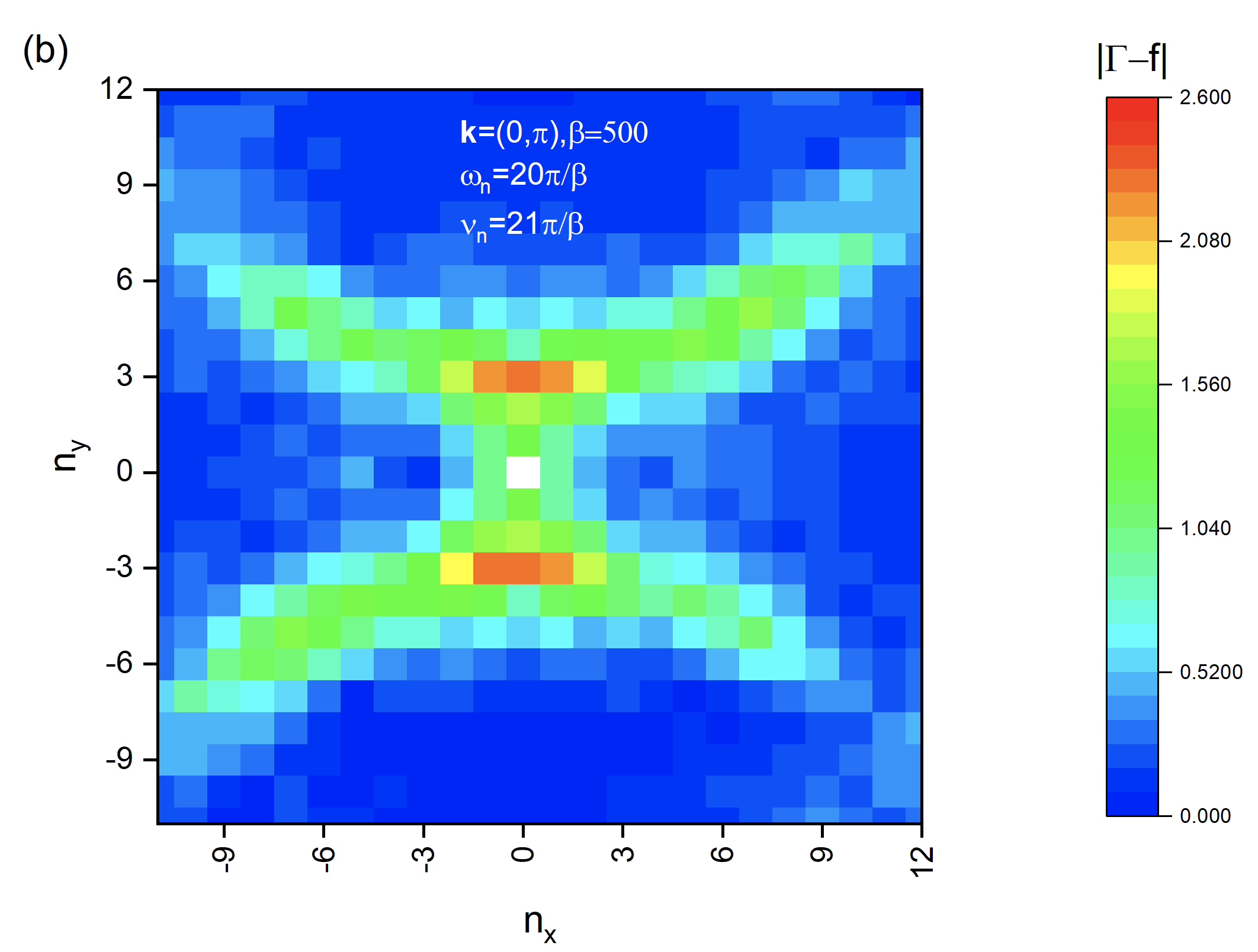}
\caption{The momentum dependence of the full vertex correction to the electron-phonon coupling from the antiferromagnetic spin fluctuation calculated in the ladder approximation. (a)The real part of the vertex correction, (b)The absolute value of the vertex correction. The calculation is done on a finite lattice with $24\times24$ sites and periodic boundary condition. The phonon momentum in the first Brillouin zone is denoted here as $\mathbf{q}=\frac{\pi}{12}(n_{x},n_{y})$, with $n_{x,y}=-11,...,12$.}
\end{figure}   
 
 \section{Intertwining effect between the electron-phonon coupling and the antiferromagnetic spin fluctuation on the electron self-energy}
The non-trivial momentum dependence of the vertex correction from the antiferromagnetic spin fluctuation to the electron-phonon coupling found in the last section implies the following counter-intuitive conclusion:  although the electron-phonon coupling to the $B_{1g}$ buckling mode is suppressed by the antiferromagnetic fluctuation in the $\mathbf{q}=0$ limit, its contribution to the PDH structure may be actually enhanced in an antiferromagnetic correlated background. In this section we present a theory for the PDH structure of the anti-nodal quasiparticle with such a vertex correction effect included. 

Since the full vertex function preserves the qualitative feature of the first order vertex correction, in the following we will consider the intertwining effect between the electron-phonon coupling and the antiferromagnetic spin fluctuation on the electron self-energy at the lowest order. The quasiparticle self-energy due to the lowest order vertex correction to the electron-phonon coupling can be represented by the two Feynman diagrams shown in Fig.13. Following the Feynman rule, the self-energy reads

\begin{equation}
\Sigma_{ver}(\mathbf{k},i\nu_{n})=\Sigma^{(1)}_{ver}(\mathbf{k},i\nu_{n})+\Sigma^{(2)}_{ver}(\mathbf{k},i\nu_{n})
\end{equation}
in which $\Sigma^{(1)}_{ver}$ and $\Sigma^{(2)}_{ver}$ denotes the self energy contribution from the first and the second diagram. They are given by 
\begin{eqnarray}
\Sigma^{(1)}_{ver}(\mathbf{k},i\nu_{n})&=&\frac{3g^{2}}{4N^{2}\beta^{2}}\sum_{\mathbf{q,q'}}f(\mathbf{k,q})f(\mathbf{k+q+q',-q})\nonumber\\
&\times&\sum_{i\omega_{n},i\omega^{'}_{n}}G^{(0)}(\mathbf{k+q'},i\nu_{n}+i\omega'_{n})\nonumber\\
&\times&\sigma_{3}G^{(0)}(\mathbf{k+q+q'},i\nu_{n}+i\omega_{n}+i\omega'_{n})\nonumber\\
&\times&G^{(0)}(\mathbf{k+q},i\nu_{n}+i\omega_{n})\sigma_{3}\nonumber\\
&\times&\chi(\mathbf{q'},i\omega'_{n})D(\mathbf{q},i\omega_{n})\nonumber\\
\end{eqnarray}
and
\begin{eqnarray}
\Sigma^{(2)}_{ver}(\mathbf{k},i\nu_{n})&=&\frac{3g^{2}}{4N^{2}\beta^{2}}\sum_{\mathbf{q,q'}}f(\mathbf{k+q',q})f(\mathbf{k+q,-q})\nonumber\\
&\times&\sum_{i\omega_{n},i\omega^{'}_{n}}\sigma_{3}G^{(0)}(\mathbf{k+q},i\nu_{n}+i\omega_{n})\nonumber\\
&\times&G^{(0)}(\mathbf{k+q+q'},i\nu_{n}+i\omega_{n}+i\omega'_{n})\sigma_{3}\nonumber\\
&\times&G^{(0)}(\mathbf{k+q'},i\nu_{n}+i\omega'_{n})\nonumber\\
&\times&\chi(\mathbf{q'},i\omega'_{n})D(\mathbf{q},i\omega_{n})\nonumber\\
\end{eqnarray}
with
\begin{equation}
G^{(0)}(\mathbf{k},i\nu_{n})=\frac{1}{i\nu_{n}\sigma_{0}-\epsilon_{\mathbf{k}}\sigma_{3}-\Delta_{\mathbf{k}}\sigma_{1}}\nonumber
\end{equation}
and
\begin{equation}
D(\mathbf{q},i\omega_{n})=\frac{-2\Omega}{\omega^{2}_{n}+\Omega^{2}}\nonumber
\end{equation}
$\chi(\mathbf{q},i\omega_{n})$ is the MMP susceptibility of the local moment fluctuation, whose spectral representation is given by
\begin{equation}
\chi(\mathbf{q},i\omega_{n})=\frac{1}{2\pi}\int^{\infty}_{0}d\omega R(\mathbf{q},\omega)\frac{-2\omega}{\omega^{2}_{n}+\omega^{2}}\nonumber
\end{equation}
Here $R(\mathbf{q},\omega)$ is the spectral weight of the local moment fluctuation.

 \begin{figure}
\includegraphics[width=8cm]{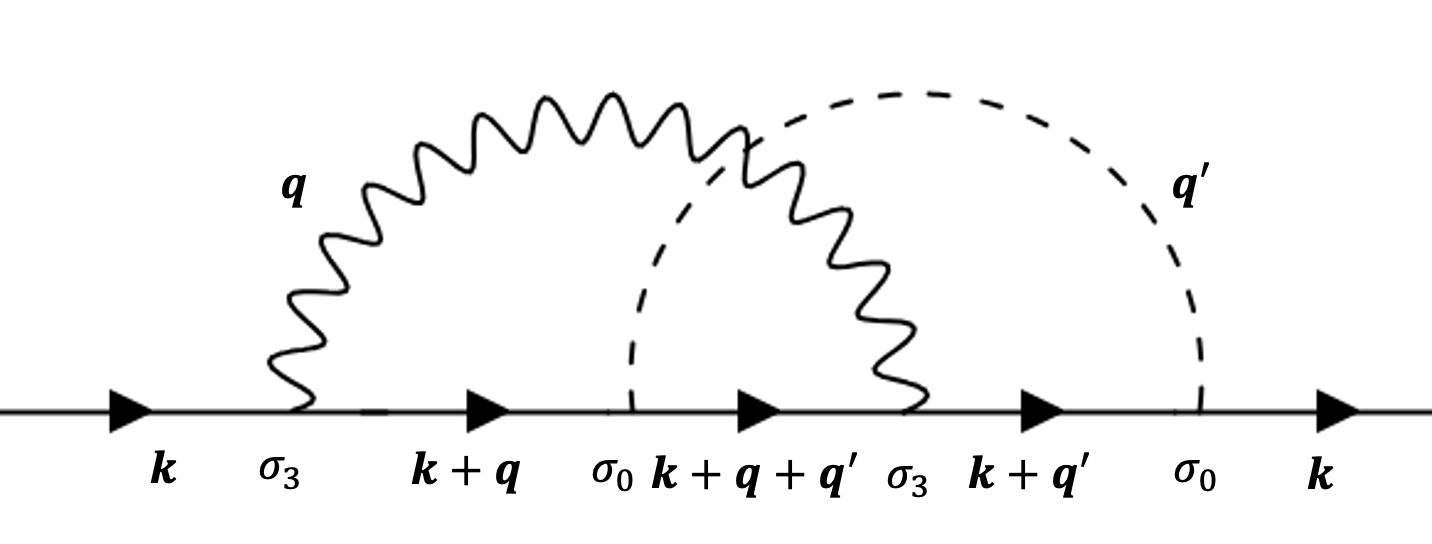}
\includegraphics[width=8cm]{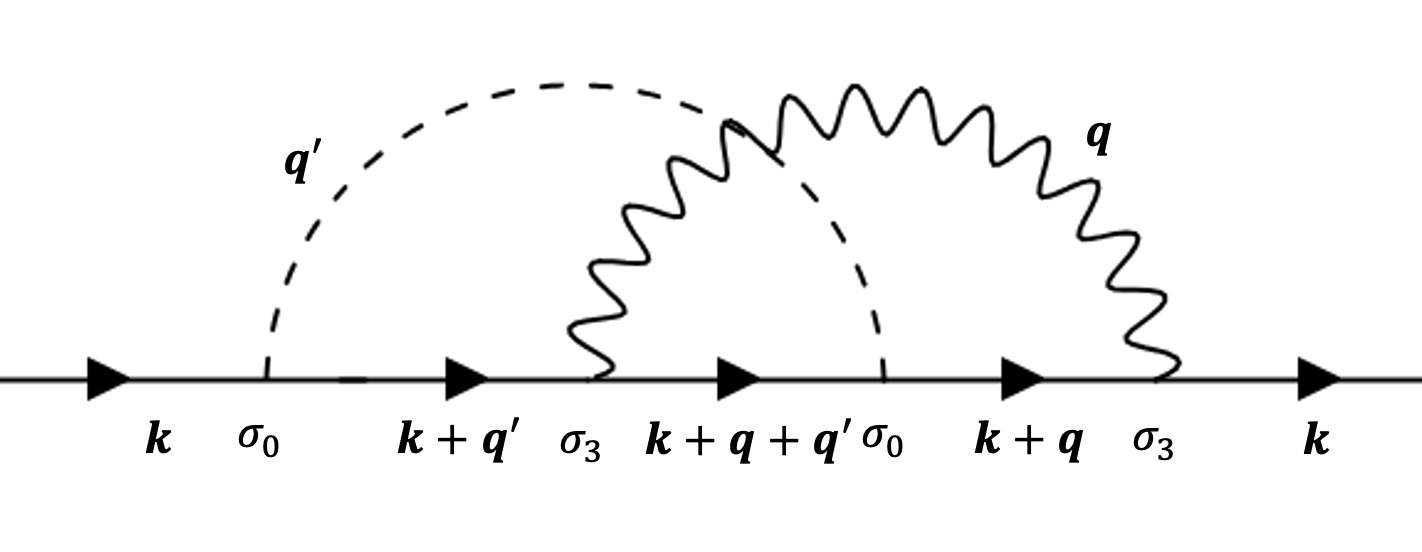}
\caption{Illustration of the intertwining effect between the electron-phonon coupling and the antiferromagnetic spin fluctuation on the electron self-energy. Shown here is the lowest order vertex correction to the electron self-energy. Here the solid lines denote the electron Green's function, the dashed line denotes the propagator of the antiferromagnetic spin fluctuation, and the wavy line denotes the phonon propagator. The upper and the lower panel denote the two conjugate scattering processes of the intertwining contribution, namely $\Sigma^{(1)}_{ver}$ and $\Sigma^{(2)}_{ver}$ in Eq.60.}
\end{figure}

Inserting the expression of $G^{(0)}$, $D$ and $\chi$ into Eq.60 we obtain
\begin{eqnarray}
\Sigma_{ver}(\mathbf{k},i\nu_{n})&=&\int_{0}^{\infty}d\omega \sum_{\mathbf{q,q'}}\left[F^{(1)}_{\mathbf{k,q,q'}}(\omega)K^{(1)}_{\mathbf{k,q,q'}}(\omega,i\nu_{n})\right.\nonumber\\
&+&\left. F^{(2)}_{\mathbf{k,q,q'}}(\omega)K^{(2)}_{\mathbf{k,q,q'}}(\omega,i\nu_{n})\right]\nonumber\\
\end{eqnarray}
in which 
\begin{eqnarray}
F^{(1)}_{\mathbf{k,q,q'}}(\omega)&=&\frac{3g^{2}}{8\pi N^{2}}f(\mathbf{k,q})f(\mathbf{k+q+q',-q})R(\mathbf{q'},\omega)\nonumber\\
F^{(2)}_{\mathbf{k,q,q'}}(\omega)&=&\frac{3g^{2}}{8\pi N^{2}}f(\mathbf{k+q',q})f(\mathbf{k+q,-q})R(\mathbf{q'},\omega)\nonumber\\
\end{eqnarray}
and
\begin{eqnarray}
K^{(1)}_{\mathbf{k,q,q'}}(\omega,i\nu_{n})&=&\frac{1}{\beta^{2}}\sum_{i\omega_{n},i\omega^{'}_{n}}G^{(0)}(\mathbf{k+q'},i\nu_{n}+i\omega'_{n})\nonumber\\
&\times&\sigma_{3}G^{(0)}(\mathbf{k+q+q'},i\nu_{n}+i\omega_{n}+i\omega'_{n})\nonumber\\
&\times&G^{(0)}(\mathbf{k+q},i\nu_{n}+i\omega_{n})\sigma_{3}\nonumber\\
&\times&\frac{2\omega}{\omega'^{2}_{n}+\omega^{2}}\frac{2\Omega}{\omega^{2}_{n}+\Omega^{2}}\nonumber\\
K^{(2)}_{\mathbf{k,q,q'}}(\omega,i\nu_{n})&=&\frac{1}{\beta^{2}}\sum_{i\omega_{n},i\omega^{'}_{n}}\sigma_{3}G^{(0)}(\mathbf{k+q},i\nu_{n}+i\omega_{n})\nonumber\\
&\times&G^{(0)}(\mathbf{k+q+q'},i\nu_{n}+i\omega_{n}+i\omega'_{n})\sigma_{3}\nonumber\\
&\times&G^{(0)}(\mathbf{k+q'},i\nu_{n}+i\omega'_{n})\nonumber\\
&\times&\frac{2\omega}{\omega'^{2}_{n}+\omega^{2}}\frac{2\Omega}{\omega^{2}_{n}+\Omega^{2}}\nonumber\\
\end{eqnarray}

After a lengthy but straightforward derivation, we find
\begin{eqnarray}
K^{(1)}_{\mathbf{k,q,q'}}(\omega,i\nu_{n})&=&A\sigma_{0}+B\sigma_{3}+C\sigma_{1}+iD\sigma_{2}\nonumber\\
K^{(2)}_{\mathbf{k,q,q'}}(\omega,i\nu_{n})&=&A\sigma_{0}+B\sigma_{3}+C\sigma_{1}-iD\sigma_{2}\nonumber\\
\end{eqnarray}
from which it can be shown straightforwardly that the $i\sigma_{2}$ component of $\Sigma_{ver}$ vanishes identically, as it should be from the time reversal symmetry of the system.
The four coefficients in $K^{(1)}$ and $K^{(2)}$ are given by 
\begin{eqnarray}
A&=&\frac{1}{2}\sum_{\{s\}}s_{1}s_{5} [\ \frac{1}{4}+(u_{2}u_{3}+v_{2}v_{3}) \nonumber\\
&+& (u_{2}+u_{3})u_{4}-(v_{2}+v_{3})v_{4}\ ] h(\{s\})\nonumber\\
B&=&\sum_{\{s\}}s_{1}s_{5}[ \ \frac{1}{4}(u_{2}+u_{3}+u_{4})+u_{2}u_{3}u_{4}\nonumber\\
&+& v_{3}(u_{2}v_{4}+v_{2}u_{4})-v_{2}u_{3}v_{4}\ ]  h(\{s\})\nonumber\\
C&=&\sum_{\{s\}}s_{1}s_{5}[\ \frac{1}{4}(v_{4}-v_{2}-v_{3})+v_{2}v_{3}v_{4}\nonumber\\
&+& u_{3}(u_{2}v_{4}+v_{2}u_{4})-u_{2}v_{3}u_{4}\ ] h(\{s\})\nonumber\\
D&=&\frac{1}{2}\sum_{\{s\}}s_{1}s_{5}[ \ (v_{2}u_{3}-u_{2}v_{3})\nonumber\\
&-& (v_{2}+v_{3})u_{4}-(u_{2}+u_{3})v_{4}\ ] h(\{s\})\nonumber\\
\end{eqnarray}
here $s_{1},s_{2},s_{3},s_{4},s_{5}=\pm1$,
\begin{eqnarray}
u_{i}&=&s_{i}\frac{\epsilon_{i}}{2E_{i}}\nonumber\\
v_{i}&=&s_{i}\frac{\Delta_{i}}{2E_{i}}
\end{eqnarray}
for $i=2,3,4$, in which we have defined the abbreviation
\begin{eqnarray}
\epsilon_{2}&=&\epsilon_{\mathbf{k+q+q'}}\nonumber\\
\epsilon_{3}&=&\epsilon_{\mathbf{k+q}}\nonumber\\
\epsilon_{4}&=&\epsilon_{\mathbf{k+q'}}\nonumber\\
\end{eqnarray}
\begin{eqnarray}
\Delta_{2}&=&\Delta_{\mathbf{k+q+q'}}\nonumber\\
\Delta_{3}&=&\Delta_{\mathbf{k+q}}\nonumber\\
\Delta_{4}&=&\Delta_{\mathbf{k+q'}}\nonumber\\
\end{eqnarray}
and $E_{i}=\sqrt{\epsilon_{i}^{2}+\Delta_{i}^{2}}$ for $i=2,3,4$. $h(\{s\})$ is defined as the following double frequency sum
\begin{eqnarray}
h(\{s\})&=&\frac{1}{\beta^{2}}\sum_{i\omega_{n},i\omega^{'}_{n}}\frac{1}{i\nu_{n}+i\omega'_{n}-s_{4}E_{4}}\nonumber\\
&\times&\frac{1}{i\nu_{n}+i\omega_{n}+i\omega'_{n}-s_{2}E_{2}}\times \frac{1}{i\nu_{n}+i\omega_{n}-s_{3}E_{3}}\nonumber\\
&\times&\frac{1}{i\omega'_{n}-s_{1}\omega}\times \frac{1}{i\omega_{n}-s_{5}\Omega}\nonumber\\
\end{eqnarray}
After some algebra one find
\begin{eqnarray}
h(\{s\})&=&(f_{2}-f_{3})n_{23}\eta_{4}\eta_{5}\eta_{6}\nonumber\\
&-&(f_{2}+n_{5})f_{25}\eta_{2}\eta_{3}\eta_{4}\nonumber\\
&+&[n_{5}f_{4}\eta_{2}+n_{1}f_{3}\eta_{6}]\eta_{1}\eta_{4}\nonumber\\
&+&[n_{5}\eta_{4}-f_{2}\eta_{6}]n_{1}\eta_{1}\eta_{3}\nonumber\\
&+&[f_{2}\eta_{2}-f_{3}\eta_{4}]f_{4}\eta_{1}\eta_{5}\nonumber\\
\end{eqnarray}
here
\begin{eqnarray}
n_{1}&=&n_{B}(s_{1}\omega)\nonumber\\
n_{5}&=&n_{B}(s_{5}\Omega)\nonumber\\
n_{23}&=&n_{B}(s_{2}E_{4}-s_{3}E_{3})\nonumber\\
f_{2}&=&f(s_{2}E_{2})\nonumber\\
f_{3}&=&f(s_{3}E_{3})\nonumber\\
f_{4}&=&f(s_{4}E_{4})\nonumber\\
f_{25}&=&f(s_{2}E_{2}-s_{5}\Omega)\nonumber\\
\end{eqnarray}
and
\begin{eqnarray}
\eta_{1}&=&\frac{1}{i\nu_{n}+s_{1}\omega-s_{4}E_{4}}\nonumber\\
\eta_{2}&=&\frac{1}{s_{4}E_{4}+s_{5}\Omega-s_{2}E_{2}}\nonumber\\
\eta_{3}&=&\frac{1}{i\nu_{n}+s_{1}\omega+s_{5}\Omega-s_{2}E_{2}}\nonumber\\
\eta_{4}&=&\frac{1}{i\nu_{n}+s_{5}\Omega-s_{3}E_{3}}\nonumber\\
\eta_{5}&=&\frac{1}{i\nu_{n}+s_{2}E_{2}-s_{3}E_{3}-s_{4}E_{4}}\nonumber\\
\eta_{6}&=&\frac{1}{s_{3}E_{3}+s_{1}\omega-s_{2}E_{2}}\nonumber\\
\end{eqnarray}

The electron self-energy with the lowest order vertex correction is thus given by
\begin{equation}
\Sigma(\mathbf{k},i\nu_{n})=\Sigma_{AF}(\mathbf{k},i\nu_{n})+\Sigma_{PH}(\mathbf{k},i\nu_{n})+\Sigma_{ver}(\mathbf{k},i\nu_{n})\nonumber
\end{equation}
The electron spectral function can be calculated from
\begin{eqnarray}
A(\mathbf{k},\omega)&=&-2Im G_{1,1}(\mathbf{k},\omega+i0^{+})\nonumber\\
&=&-2Im[[G^{(0)}(\mathbf{k},\omega+i0^{+})]^{-1}-\Sigma(\mathbf{k},\omega+i0^{+})]^{-1}_{1,1}\nonumber
\end{eqnarray}

A numerical calculation of the electron spectral function from the above formula is straightforward. However, since there are large uncertainties in the value of the coupling constant $g$ and $\lambda$ and that the magnitude of the vertex correction $\Sigma_{ver}$ depends sensitively on these parameters, here we will restrict ourselves to a qualitative discussion. More specifically, we will consider the limiting case in which either $\Sigma_{PH}$ or $\Sigma_{AF}$ is the dominant contribution to the electron self energy and that $\Sigma_{ver}$ is a small correction to $\Sigma_{PH}$ or $\Sigma_{AF}$. In Fig.14, we plot the electron spectral function calculated at the anti-nodal point with $\Sigma(\mathbf{k},i\nu_{n})=\Sigma_{PH}+\alpha\Sigma_{ver}$, in which $\alpha$ as a tuning parameter. Here we have adopted a small value for $g$. As can be seen from the figure, the PDH structure strengthens with the increase of $\alpha$. This is consistent with the analysis we presented in the last section. In the opposite limiting case when $\Sigma_{AF}$ is the dominant contribution to the electron self energy, $\Sigma_{ver}$ can also be interpreted as the vertex correction to $\Sigma_{AF}$ from the electron-phonon coupling. In Fig.15, we plot the electron spectral function calculated at the anti-nodal point with $\Sigma(\mathbf{k},i\nu_{n})=\Sigma_{AF}+\alpha\Sigma_{ver}$, in which $\alpha$ as a tuning parameter.  Here we have adopted a small value for $\lambda$. As can be seen from the figure, the PDH structure also strengthens with the increase of $\alpha$. To sum up, the coupling with the antiferromagnetic spin fluctuation and the $B_{1g}$ buckling phonon cooperate with each other in producing the PDH structure in the anti-nodal quasiparticle spectrum. 

 \begin{figure}
\includegraphics[width=8cm]{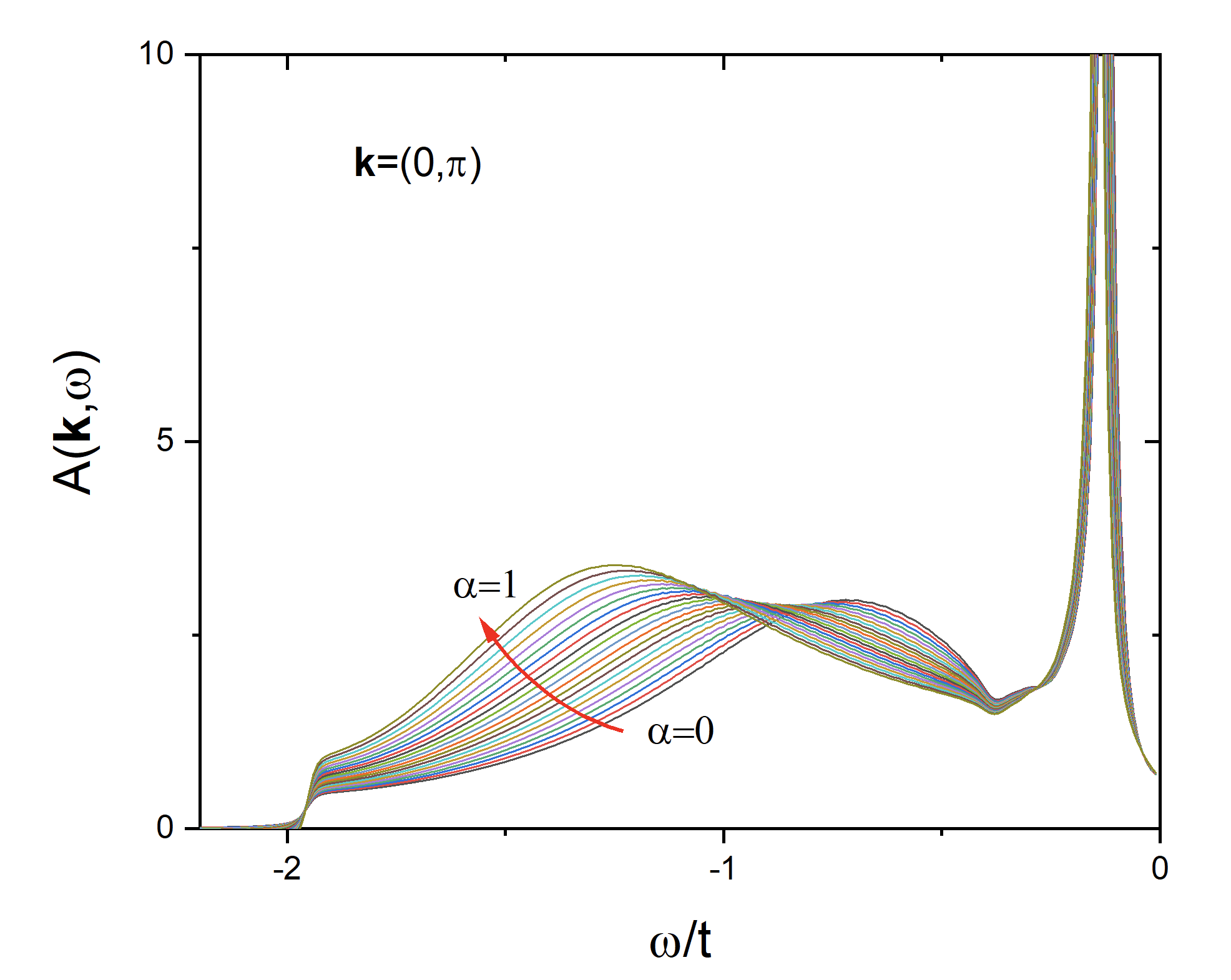}
\caption{Illustration of the intertwining effect between the electron-phonon coupling and the antiferromagnetic spin fluctuation on the electron self-energy. Plotted here is the electron spectral function calculated at $\mathbf{k}=(0,\pi)$ with $\Sigma(\mathbf{k},i\nu_{n})=\Sigma_{PH}+\alpha\Sigma_{ver}$, in which $\alpha$ as a tuning parameter. The PDH structure is seen to strengthen with the increase of $\alpha$. Thus the vertex correction from the antiferromagnetic spin fluctuation enhances the phonon contribution to the PDH structure.}
\end{figure}  

 \begin{figure}
\includegraphics[width=8cm]{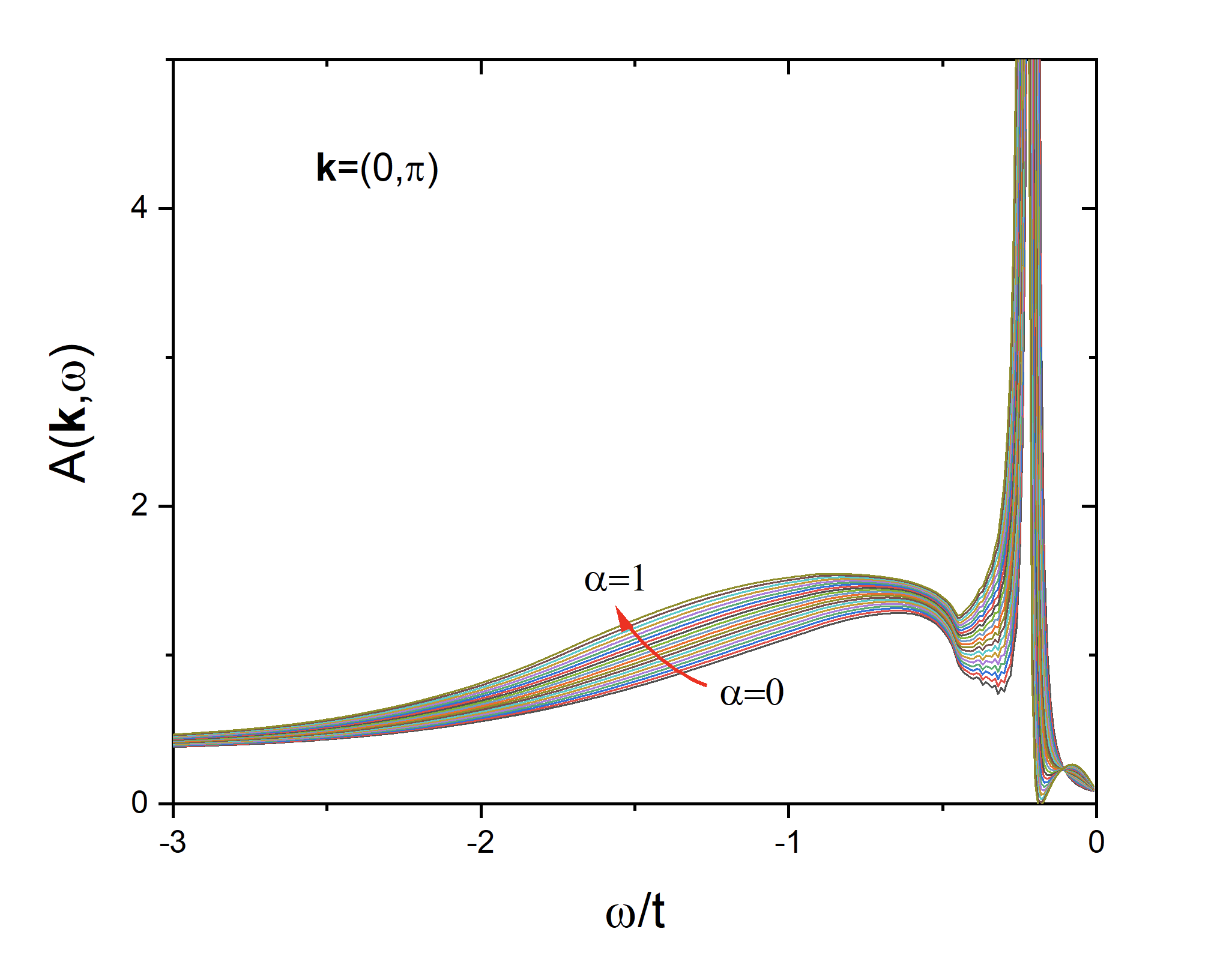}
\caption{Illustration of the intertwining effect between the electron-phonon coupling and the antiferromagnetic spin fluctuation on the electron self-energy. Plotted here is the electron spectral function calculated at $\mathbf{k}=(0,\pi)$ with $\Sigma(\mathbf{k},i\nu_{n})=\Sigma_{AF}+\alpha\Sigma_{ver}$, in which $\alpha$ as a tuning parameter. The PDH structure is seen to strengthen with the increase of $\alpha$. Thus the vertex correction from the electron-phonon coupling also enhances the spin fluctuation contribution to the PDH structure.}
\end{figure}  

\section{Conclusions and discussions}
In this work we have conducted a systematic perturbative study on the interplay between the antiferromagnetic spin fluctuation and the electron-phonon coupling in the cuprate superconductors. The focus is on their respective roles in the origin of the PDH structure of the anti-nodal spectrum. The main results of the study can be summarized as follows.

Firstly, we find that the coupling to either the antiferromagnetic spin fluctuation or the $B_{1g}$ buckling phonon mode can generate a PDH structure in the anti-nodal spectrum. In both scenarios such a spectral anomaly can be attributed to the level repulsion effect between the bare quasiparticle dispersion and that of the scattering final state. However, the momentum transfer of the dominate scattering process is very different in these two scenarios. More specifically, in the antiferromagnetic spin fluctuation scenario, the PDH structure is mainly contributed by scattering final state located around the other anti-nodal point in the Brillouin zone. On the other hand, such a spectral anomaly is mainly contributed by scattering final state located around the same anti-nodal point in the electron-phonon coupling scenario. The dominate momentum transfer is thus around $\mathbf{Q}=(\pi,\pi)$ and $\mathbf{q}=(0,0)$ respectively in these two scenarios.

Secondly, we find that the matrix element of the electron-phonon coupling to the $B_{1g}$ buckling mode has a highly non-trivial momentum structure. More specifically, we find that the electron-phonon coupling matrix element change sign when the electron momentum change by $\mathbf{Q}=(\pi,\pi)$. Such a peculiar momentum dependence plays an interesting and intriguing role in the interplay between the electron-phonon coupling and the antiferromagnetic spin fluctuation in the system. Most importantly, the coupling strength to the $B_{1g}$ buckling mode suffers from a destructive interference effect in an antiferromagnetic correlated background. For example, in an antiferromagnetic long range ordered state, the intra-band matrix element of the electron-phonon coupling in the SDW-split band would be suppressed to zero at the hot spot of the underlying fermi surface. A first order perturbative calculation on the vertex correction also indicates that the coupling strength to the $B_{1g}$ buckling mode is strongly suppressed by the antiferromagnetic spin fluctuation in the $\mathbf{q}=0$ limit. In real space, such a destructive interference effect can be understood as the consequence of the competition in low energy electron spectral weight between two channels, namely the antiferromagnetic spin correlation between NN sites and the NN hopping term that the $B_{1g}$ buckling mode couples to. Naively, one would thus conclude that the phonon contribution to the PDH structure should be suppressed in a background with strong antiferromagnetic spin fluctuation as a result of such a destructive interference effect.   

Surprisingly, we find that such a naive expectation is false. More specifically, we find that the vertex correction from antiferromagnetic spin fluctuation is strongly anisotropic in the momentum space around the $\mathbf{q}=(0,0)$ point for the anti-nodal quasiparticles. Such a strong anisotropy in the vertex correction can be attributed to the strong anisotropy in the quasiparticle dispersion around the anti-nodal point, which is a saddle point of the band dispersion. A direct consequence of such a non-trivial momentum dependence is the strong enhancement of the electron-phonon coupling strength in two crescent-shaped regions around the $\mathbf{q}=(0,0)$ point. Intriguingly, this is just the momentum region where the $B_{1g}$ phonon mode contribute the most to the PDH structure. We find that such a peculiar momentum structure in the vertex correction is preserved in the full vertex function at least under the ladder approximation. We thus expect that the phonon contribution to the PDH structure to be enhanced, rather than suppressed by the vertex correction from the antiferromagnetic spin fluctuation. This expectation is confirmed by our direct calculation of the electron spectral function with such vertex correction effect included.

Thus, contrary to the naive expectation, the coupling to the antiferromagnetic spin fluctuation and the $B_{1g}$ buckling mode actually act cooperatively in the generation of the PDH structure in the anti-nodal spectrum. Such strong intertwinement between the antiferromagnetic spin fluctuation and the electron-phonon coupling makes it even more subtle to decipher the true implication of the PDH structure for the cuprate physics. The recently observed sudden suppression of the PDH structure around the pseudogap end point may provide the key clue to resolve this mystery. In a phonon dominated scenario, such a sudden suppression of the PDH structure would imply a corresponding sudden suppression of the bare coupling strength to the $B_{1g}$ buckling mode. While this is indeed possible, it is hard to understand why such a sudden suppression occurs right at the pseudogap end point, since the majority of experimental evidences indicate that the electron-phonon coupling plays at most a secondary role in the origin of the pseudogap phenomena in the cuprate superconductors.         

The situation becomes much more transparent if we assume a spin fluctuation dominated scenario. Upon entering the pseudogap phase from the over-doped side of the cuprate phase diagram, strong antiferromagnetic fluctuation of local moment will emerge in the system. More accurately, there will be a sudden transmutation in the nature of spin fluctuation from itinerant particle-hole-continuum-like to local-moment-like around such a critical doping, as is confirmed by recent RIXS measurement. Such a change will be accompanied by the emergence of the PDH structure in the anti-nodal spectrum, which can be attributed to the coupling between the quasiparticle and the antiferromagnetic fluctuation of the emergent local moment in such a doped Mott insulating state. More quantitatively, the rapid rise in the strength of the PDH structure around the pseudogap end point can be attributed to both the enhancement of the antiferromagnetic spin fluctuation and the accompanying enhancement in the phonon contribution to the PDH structure through the vertex correction by the same emergent antiferromagnetic spin fluctuation. 

According to this picture, the sudden emergence of the PDH structure in the anti-nodal spectrum should be understood as a spectral signature of the emergence of fluctuating local moment or the entrance of the doped Mott insulating state. There is no need to assume sudden change in the bare electron-phonon coupling strength around the pseudogap end point. Such an understanding is consistent with the result of recent attempt to measure the relative strength of the electron-phonon coupling in the cuprate superconductors with the RIXS technique. More specifically, it is found that the doping dependence of the electron-phonon coupling in the cuprate superconductors around the pseudogap end point is much more smooth than that needed to account for the sudden suppression of the PDH structure in the phonon scenario. In fact, the electron-phonon coupling strength estimated from the RIXS measurement can even exhibit non-monotonic doping dependence around the pseudogap end point for small $\mathbf{q}$\cite{Peng}.

Finally, we see from this work that the momentum dependence of the electron-phonon coupling matrix element can play a vital in a strongly correlated background, although the lattice vibration related to the phonon mode may itself be local. A through understanding of such non-trivial momentum dependence may shed important light to many mysteries of the cuprate physics. In particular, the interplay between the electron-phonon coupling with such non-trivial momentum dependence and the Mott physics may play an intriguing role in both the spin, charge and the pairing channel. The perturbative treatment of such intertwinement effect as presented in this work should in general be replaced by more sophisticated and more quantitative non-perturbative treatment in these situations. We leave the study of these issues to future endeavor.

 \begin{acknowledgments}
We acknowledge the support from the National Natural Science Foundation of China (Grant No. 12274457).
\end{acknowledgments}

\end{document}